# Polymer Matrix Sand Composites for Enhanced Ballistic Impact Resistance

Manas Thakur[*], Nakka Nishika[ǂ], Bommiditha Jyothsnavi[ǂ], Surkanti Sai Sahasra[ǂ], Shiva Bansal[#], Rajendra Kumar Munian[ǂǂǂ], Srikant Sekhar Padhee[ǂǂǂǂ]

[*] Dept. of Mechanical Engineering, MTech Student, Indian Institute of Technology Ropar, Rupnagar, India
[ǂ] Dept. of Mechanical Engineering, B.Tech. Student, Indian Institute of Technology Ropar, Rupnagar, India
[#] Dept. of Mechanical Engineering, PhD Scholar, Indian Institute of Technology Ropar, Rupnagar, India
[ǂǂǂ] Dept. of Mechanical Engineering, Assistant Professor, Indian Institute of Technology Ropar, Rupnagar, India
[ǂǂǂǂ] Dept. of Mechanical Engineering, Associate Professor, Indian Institute of Technology Ropar, Rupnagar, India

Email: *manasthakur159@gmail.com*
*nishika1028@gmail.com*
*jyothsnavi987@gmail.com*
*saisahasrasurkanti@gmail.com*
*shiva.19mez0011@iitrpr.ac.in*
*rajendra.munian@iitrpr.ac.in*
*sspadhee@iitrpr.ac.in*

**Abstract.** With the ever-increasing threat of ballistic impact, it is essential to provide a solution that is not only effective but also economical. A majority of studies contribute toward alternatives to monolithic structures by incorporating sandwiched cores, which are often prone to core crushing and delamination. This often limits the multi-hit capabilities of the structure. In recent years, sand-based composites have emerged as a potentially cost-effective solution. In this ongoing effort, the current investigation aims to offer more robust protection against varied ballistic impacts and potential ballistic threats. This research investigates the enhancement of ballistic impact resistance in Polymer Matrix Sand Composites (PMSCs) through the inclusion of sand in a graded fashion, resulting in a tunable solution. The tailoring of various mechanical properties, such as modulus, impact strength, and hardness, enables different layers within a single structure, offering potential advantages as the projectile pierces through the thickness. The gradation creates a stepwise structure, with a prime base impact zone densely graded with inclusions that are brittle and hard enough to erode the projectile. Further gradation involves a less dense region providing tensile strength, which can reflect the tensile wave and reduce impact energy. In the preceding study, composites were fabricated and subjected to a battery of tests, including tensile testing, Izod impact testing, and hardness testing, to comprehensively evaluate their mechanical and physical properties also Homogenized properties were extracted so as to critically observe the PMSCs behavior upon variations in constituent elements. In the current study, the properties extracted from variations in constituent elements are utilized for developing different layers in a structure. Simulation studies of varied gradation configurations are conducted, and the potential advantages of layering sequences are studied. Preliminary projectile impact studies provide initial insights into gradation performance. The experimental and simulation results reveal that varying the size and volume fraction of sand content markedly influences the mechanical properties and ballistic resistance of the composites, underscoring the potential of PMSCs as cost-effective and environmentally sustainable Composites for advanced ballistic protection applications.

# 1 Introduction

Effective ballistic-resistant structures often exhibit an amalgamation of several layers with varied material configurations or distinguished structural configurations. This combination ultimately provides an integration of layers with high hardness, toughness, brittleness, lightweight, and high-energy absorbent characteristics. Guiding upon the potential of monolithic structures [1] often comprising a single material configuration such as iron steels or aluminum alloys for potential ballistic-resistant structures, their aerial densities result in weight ineffectiveness [2]. Additionally, their often low multi-hit capabilities and overall cost-ineffectiveness make these structures limitedly suitable for ballistic impact applications [3]. Studies have been carried out to reduce the weight of monolithic structures [5] by incorporating sandwich structures comprising regions with significant characteristics that work together as the bullet pierces through the structure's thickness. Typical configurations for ballistic impacts include a frontal region often made of Rolled Homogeneous Armor (RHA) steel [7] or harder ceramic plates, which absorb the initial impact, deteriorate the bullet tip, and resist further penetration by increasing the contact area, thereby mitigating localized stresses and increasing penetration time [6]. Frontal brittle and hard materials like ceramics dissipate most of a bullet's kinetic energy through crack formation and fragmentation, often failing catastrophically upon multi-hit impacts. To address this, energy-absorbent structures like Composite Metal Foams (CMFs) and corrugated layers are integrated as middle layers to support the ceramics, reduce weight, and distribute energy over a larger area [7]. These designs enhance ballistic limits with lower areal densities compared to monolithic structures [8-10].

Along with the incorporation of lightweight energy absorbent structures Lightweight Fiber reinforced composites are employed due to the exceptional capabilities of composite materials to provide superior protection while maintaining the lightweight profile. The composite materials provide excellent strength and can dissipate kinetic energy over a larger area by incorporating effective stacking sequences [11]. Also, the tailoring ability of the composite materials by virtue of their constituent elements i.e. fiber, matrix, inclusions, etc. makes them suitable to be used in a variety of applications [12]. High strength/modulus fibers such as glass fibers, ceramic fibers, carbon fiber, Para-Aramid fibers, UHMWPE fibers [4], Aromatic polyester, etc. are being used widely in the studies confined with fiber reinforced ballistic resistant structures [13-15]. Fiber-reinforced composites composed of synthetic fibers, such as aramid or carbon, offer enhanced ballistic resistance when integrated into sandwich structures due to their superior energy dissipation capabilities. The fiber layers absorb and distribute impact forces efficiently, while the sandwich core (typically foam or honeycomb) further mitigates damage by delocalizing stresses and reducing local deformation. This combination provides improved multi-hit resistance, reduced back-face deformation, and enhanced stiffness-to-weight ratio, optimizing the composite's performance against high-velocity projectiles [16-18]. Leveraging the capabilities of synthetic fiber composites, Natural fibers have also been studied extensively in ballistic applications for their easy availability, eco-friendly, and economical alternatives to synthetic fiber composites. Flax, hemp, and jute fibers are commonly studied natural fibers that offer potential in ballistic applications due to their biodegradability, low density, and cost-effectiveness [19-20]. When used in fiber-reinforced composites, these fibers contribute to ballistic resistance by absorbing impact energy and delaying projectile penetration. However, their relatively lower tensile strength and stiffness compared to synthetic fibers limit their standalone effectiveness. To enhance their performance in ballistic applications, these natural fibers are often combined with synthetic fibers in hybrid composites [20], where they contribute to energy dissipation while the synthetic fibers handle higher loads and impact forces [21-22]. Additionally, surface treatments and resin systems are employed to improve fiber-matrix adhesion, further optimizing the ballistic efficiency of these natural fiber composites. The ability of Functionally graded materials to provide a novel class of materials in which thermo-mechanical properties can be varied in three directions has been extensively studied in the area of projectile impact protection [23-24]. The inherent smooth material properties variations in a single structure have provided the ability to produce a structure with necessary directional gradients according to the requirements [25].

Summing upon the exceptional ballistic impact resistance of the different structures, it is also crucial to analyze the potential issues that are adhered to the configurations. In sandwich structures where primarily, the initial layers are made of brittle materials leads to abrupt collapse of the structure resisting its multi-hit capability, often requires strong backing for fragmentation, and is comparatively Heavy and Cost-ineffective [7]. The major issues lie within the energy absorbent structures i.e. CMFs, Corrugated Sheets, etc. their Manufacturing complexity, and Low Multi-Hit Capabilities due to deformed cells or corrugations [8]. Also, the Density w. r. t. energy absorption of the structures is difficult to attain. The overall sandwich structures face delamination of the regions which is a major concern [10]. In Polymer Matrix Fiber reinforced composites where the matrix is generally brittle in nature inhibit matrix cracking which leads to degradation of the structural stability. Delamination between the laminae and fiber pullout is one of the major drawbacks concerning the mitigation of failure [14]. Fibers such as aramid which are generally used in ballistic applications are typically moisture sensitive and degrade upon extreme weather conditions. Additionally, manufacturing complexity and Cost-ineffectiveness are also major concerns regarding the fiber-reinforced composite Structures [15].

Inclusion composites, which incorporate various fillers like sand, ceramics, or metallic particles into a polymer matrix, offer significant potential in ballistic applications. The embedded inclusions enhance the material's ability to absorb and distribute impact energy, contributing to improved resistance to high-velocity projectiles. In a polymer matrix, these composites can be engineered to create localized zones of higher hardness and toughness, increasing the system's overall durability [40]. When used in armor systems or protective panels, the combination of energy dissipation from the polymer and the enhanced mechanical properties from the inclusions can effectively reduce damage from ballistic impacts. Several studies have been carried out to extract the effective properties of inclusions of different sizes, shapes, and aspect ratios [26-30]. Sand-filled composites are the great contributors to enhance the ballistic impact resistance. These composites typically incorporate sand as a filler to improve hardness and stiffness, leveraging the availability and low cost of sand. However, the impact resistance of these composites is often suboptimal due to the lack of a systematic approach to selecting the type, size, and distribution of sand particles, as well as the methods used for mixing and curing the composites. The dilation of the overall composite with sand inclusion helps in impact resistance and penetration due to lower void spaces in between the sand grains which does not make the particles rearrange them to more dense condition. The relative density of sand also affects the penetration and erosion to some extent also projectile shape and mass along with sand's relative density affect the Ballistic limit of the sand [31]. Several researchers examined the enhancement of matrix by incorporating rubber and sand as a filler material and studied the effect of vulcanization on composite material properties which showed an increase in energy absorption of composite materials due to their increased ductility [32-33]. The major properties of the composites formed are proportional to the appropriate manufacturing and fabrication process adopted which includes sand preparation, sand segregation, sand treatment, sand mixing, etc. Also, incorporating the appropriate pre-curing and post-curing procedures enhances the properties of the overall media up to a certain extent [34-40].

In the preceding research work, the influence of sand inclusions in two compositions of the polymer matrix forming a composite was studied. The fabrication process of the composite structures was refined to obtain optimum mechanical properties by varying the sand inclusion size and weight fraction in the polymer matrix. A battery of tests has been conducted to extract experimental mechanical properties, including the uniaxial tensile test, Izod impact test, Shore-D hardness test, etc. The property extraction samples were prepared by improving the fabrication process under each distinct condition. Simulation studies were carried out to extract the effective properties of the Representative Volume Element (RVE) by varying the inclusion size and volume fraction. The current work incorporates simulation studies to observe the effect of different layering sequences and layers incorporated in a single structure, achieved by employing the experimentally extracted material properties. The layering or gradation sequence is arranged to obtain variable mechanical properties in a single structure through the tailoring ability of PMSCs due to their constituent elements, i.e., sand inclusions and the polymer matrix. The multilayered structure has initial layers densely graded with sand inclusions, responsible for initial impact absorption and abrasive tip deterioration of the bullet, which helps mitigate localized stresses and increases the contact area, leading to prolonged bullet contact time with the structure. The subsequent, less dense region provides necessary support to the frontal layer and absorbs residual energy, transferring stresses over a larger area. The final, barely dense, or neat matrix region prevents blackface signature due to the bullet's momentum and provides a cushioning effect. Simulation studies for gradation within a single structure and for the whole structure using a single material were also conducted against various projectile velocities to extract the energy absorption of each layer and determine critical velocities for specific configurations. The loss in the K.E. of the bullet was observed for each configuration and thereby the ballistic limits were obtained. The composites exhibit superior interfacial bonding between the matrix and fillers, preventing delamination under high-velocity impacts. This structural integrity ensures that the layers remain intact, even under extreme stress, thereby enhancing the material's overall ballistic resistance and energy absorption capabilities.

## 2   Simulation

### 2.1   Effective Properties Evaluation

In order to analyze the effect of inclusion size and volume fraction on the macro-scale DIGIMAT FEA was used and effective properties were extracted from the RVE. The variation was carried out w. r. t. the inclusion size and volume fraction in an RVE where inclusions were considered spherical in nature considering the random but uniform distribution of inclusions over the RVE matrix media. The sand inclusion size was varied from 0.1 mm to 1 mm in 0.1mm steps in each RVE with a variation of inclusion volume fraction from 0.1 to 0.6 in most of the simulations in the RVE media as illustrated in **Figure 2.**

Displacement Boundary condition was imposed to extract different effective properties as a negligible variance was observed in the effective properties imposed by other boundary conditions [55]. Effective Properties were extracted by the Volume Averaging Method where the stresses and strains in the respective directions as per the boundary conditions imposed are averaged out over the volume which ultimately gives effective properties in the particular direction and plane. Some of the typical representation of Modelling RVE for varied inclusion size and volume fraction and distribution of varied inclusion size and volume fraction of inclusions over the Matrix is Shown in **Fig 7**. The properties of the constituent elements i.e. Matrix and sand were assigned as per **Table A2** and **Table A3**. The Properties of LY 556 and HY 951 combined were taken from the Manufacturer's Catalogue, and the sand properties Were assigned as per [47-49] Considering the properties for Construction sand with the most quartz-silica content.

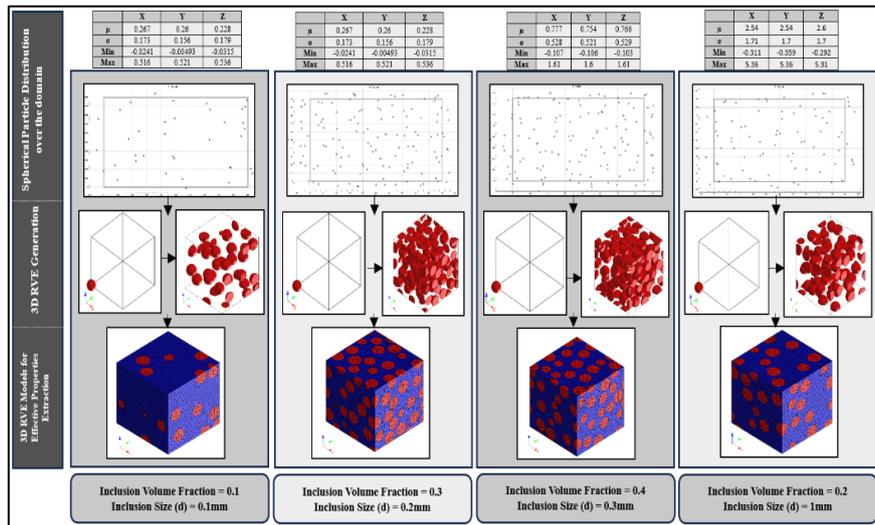

**Fig. 2.** Modelling of RVE for Varying inclusion size and volume fractions along with the distribution of inclusions over RVE

## 2.2 Computational Projectile Impact Model

In order to predict the material behavior against ballistic threats, Finite Element Models were developed and utilized by using the Commercial Abaqus/Explicit FEA tool. The material properties and distinguished damage criteria according to the change in material were assigned to the respective sections. In order to have a model that is computationally relatively intensive and accurate, proper methodologies were utilized which are narrated in the succeeding sections.

### 2.2.1 Geometry

The simulation studies feature a projectile with a hemispherical nose of radius 4 mm. The overall height of the projectile is 19 mm, with a width of 8 mm at the base. The impactor was assigned varying velocities to analyze the impact. The impacted structure is modeled with varying gradations along the thickness of the plate with varying material properties assigned to each gradation. The overall dimensions of the plate were taken as per the experimental impact analysis dimensions where the dimensions of the structure were 130 X 130 X 40 mm and 130 X 130 X 20 mm a typical model for projectile and structure is as in **Figure 1 a** and **Figure 1 b.**

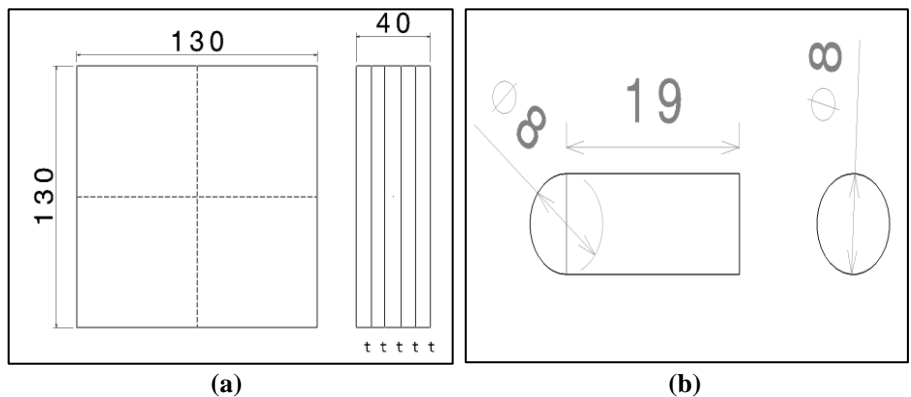

(a)            (b)
**Fig. 1.** (a) Impacted Structure (b) Hemispherical Projectile

### 2.2.2 Boundary Conditions

To optimize computational cost and enhance the effectiveness of the model, the structure was divided into four equal quadrants. Symmetric boundary conditions were imposed on the designated faces of the quarter model, effectively reducing the computation time and resource requirements while preserving the integrity and accuracy of the simulation results. This approach leverages the geometric and loading symmetry of the model, simplifying complex simulations into manageable sub-problems without loss of generality. Symmetry boundary conditions were applied along faces OA and OB, as illustrated in **Fig. 3.a)** . These conditions ensure that the modeled quarter behaves identically to the corresponding sections of the full-scale structure. By employing this method, computational efficiency is achieved by solving only one-quarter of the full problem domain, enabling the analysis to maintain high fidelity while reducing computational demand by approximately 75%. Subsequent analyses focused on validating the quarter model's response, which was extrapolated to represent the behavior of the entire structure. This process ensured that deformation patterns, stress distributions, and wave propagation characteristics were consistent across the quarter and full-scale models. The imposition of symmetry conditions also facilitated parallel computing, which further minimized simulation run times and enhanced throughput for high-resolution impact studies. This approach is particularly beneficial for simulations involving transient, non-linear dynamic responses, where the computational expense can be significant. The Boundary Conditions Imposed were as follows :
- Symmetricity-X B.C.s – Face OA ,where U1=0
- Symmetricity-Y B.C.s – Face OB, where U2=0
- ENCASTER B.C.s on the other faces, where U1=U2=U3=0

Further, summing upon the area of interest on the structure, the Middle section of the domain was divided, and a partition was created which further helps in focusing upon the driven results in the area. The Middle Part was divided into CED as depicted in **Fig.3. b**.

### 2.2.3   Meshing and Mesh Convergence

In order to obtain results driven by computational intensiveness and comparable accurate predictions, Mesh Convergence analysis was carried out in which the Divided area of focus i.e. CDE was assigned a relatively finer mesh than the other domain. A 3D reduced integration hexahedral mesh (C3D8R) was employed for the finite element analysis [50]. This 8-node brick element, with reduced integration and a single integration point, is particularly well-suited for large deformation and impact simulations. The use of C3D8R elements helps mitigate computational costs while maintaining a balance between accuracy and efficiency in capturing the deformation and stress distribution during the ballistic impact event. Additionally, hourglass control was activated to prevent spurious modes of deformation, which can arise from reduced integration. A typical representation of the meshed part is depicted in **Figure 3 .c)**. Mesh convergence analysis was performed by varying the number of elements along the thickness, evaluating the bullet's Residual Kinetic Energy (RKE) and Depth of penetration as a function of the number of elements serving as the criterion for convergence which is represented in **Fig 5**. The number of elements along the thickness of the impact region of 0.025 meter varied from 8 to 24 elements resulting in 1.66e4 to 6.78e5 elements in the overall domain. Since the computational cost with regards to 20 elements i.e. 2e5 elements is less, this mesh was selected for further analysis with approximately 32 to 33 elements along the longer boundaries to make a tradeoff between accuracy and computational time required for simulating the model. The finalized Mesh Domain is depicted in **Fig 4.**

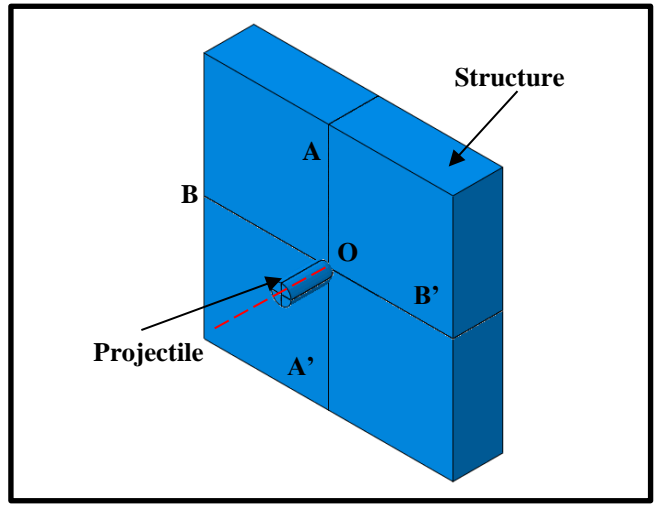

**Fig. (a)**

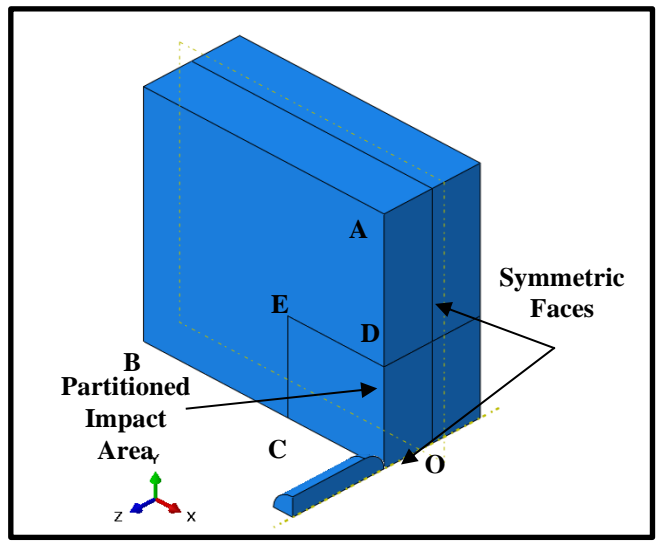

**Fig. (b)**

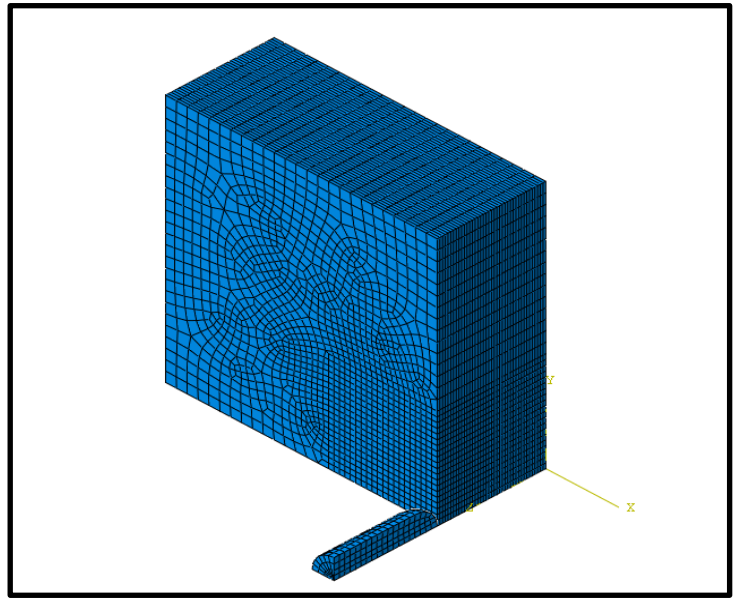

**(c)**
**Fig. 3. (a)** Overall, Domain (b) Symmetric Computational model (c) Meshed Model

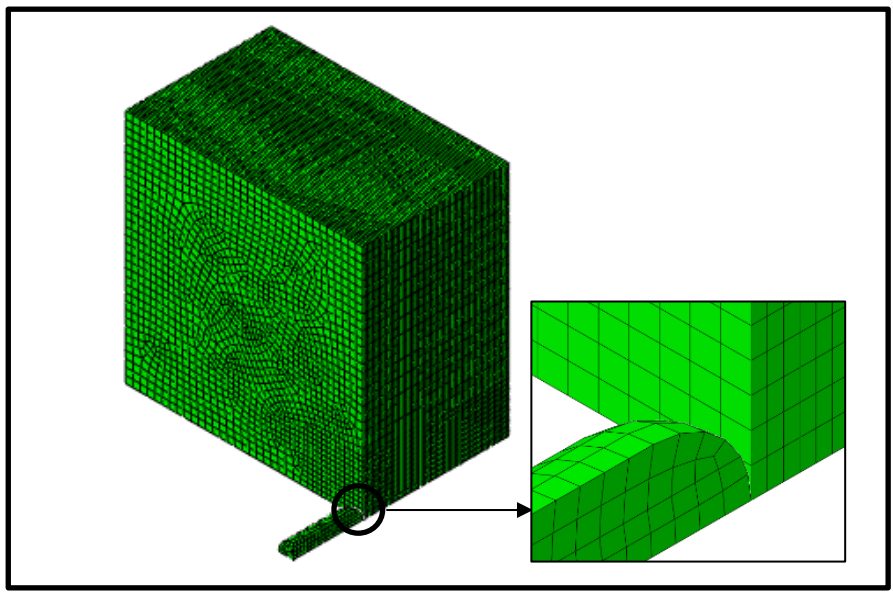

**Fig. 4.** Finalized Meshed Model with exploded view of finer mesh on area of impact

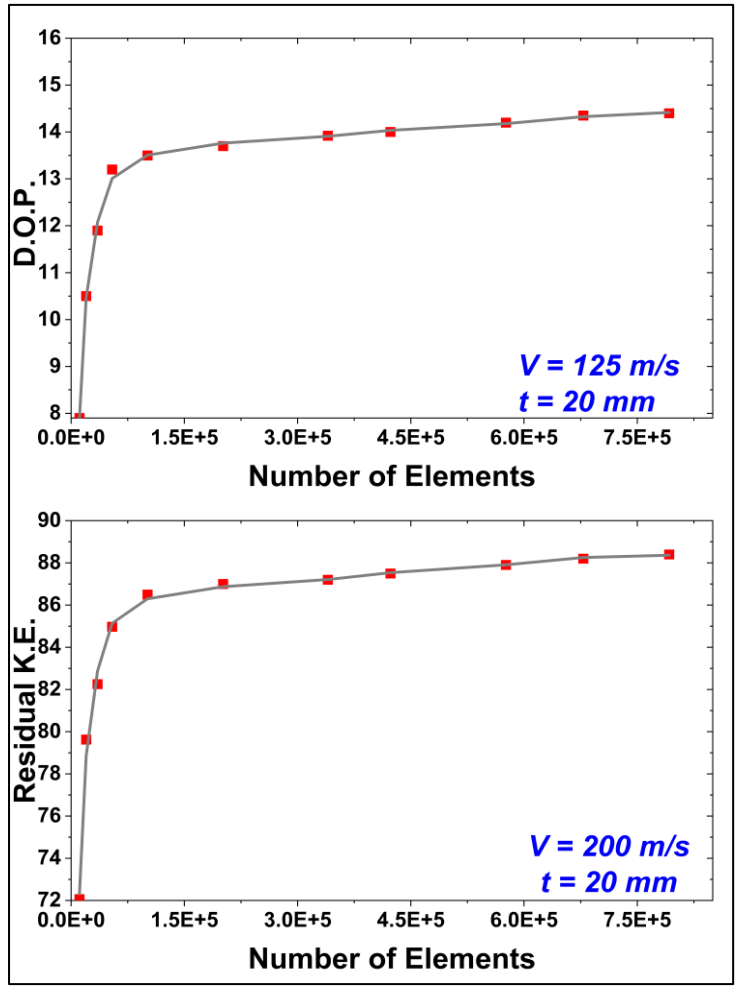

**Figure 5.** Mesh Convergence Analysis (Residual K.E. Vs No. Of elements)

### 2.2.4  Material Models and Material Assignment

Various material models, including damage and plasticity frameworks, were utilized to accurately capture material behavior under the high-strain-rate conditions of ballistic impact. These models simulate the response of materials when subjected to extreme stress, such as in the case of bullet penetration. By accounting for both damage initiation and material plasticity, the models effectively represent the material's capacity to absorb energy and resist fracture, thereby providing valuable insights into failure mechanisms during high-velocity projectile impacts.

### 2.2.4.1 Johnson-Cook plasticity model

The Johnson-Cook plasticity model is widely used for simulating the behavior of materials under high strain rates, large deformations, and varying temperatures, making it particularly effective for ballistic and impact applications [51]. This model defines the flow stress of a material as a function of strain, strain rate, and temperature, allowing for a realistic simulation of material behavior during dynamic events. One of the key advantages of the Johnson-Cook model is its ability to capture strain-hardening effects and temperature softening, which are crucial for predicting failure in high-speed impacts. It is particularly beneficial for metals and alloys, where understanding material behavior under extreme loading conditions is essential for accurate failure prediction and energy absorption. The model's simplicity and robustness make it ideal for high-velocity simulations in applications such as defense, aerospace, and automotive industries. JC Constitutive model majorly accounts for strain, strain rates, and temperature as depicted in Eq. (1)

$$\sigma_y(\varepsilon_P, \dot{\varepsilon}_P, T) = [A + B(\varepsilon_P)^n][1 + C \ln(\dot{\varepsilon}*_P)][1 - (T*)^m] \quad (1)$$

The first term in the bracket accounts for the influence of strain at the deformation, where $\varepsilon_P$ represents cumulative plastic strain. $A$ is the initial yield stress at the specific strain rate and temperature., $B$ is the hardening parameter, and n is the exponent on the equivalent plastic strain. The second term describes the effect of strain rate, where $C$ is the strain rate sensitivity constant, and $\dot{\varepsilon}*_P$ is the ratio of the current strain rate $\dot{\varepsilon}_P$ to a reference strain rate $\dot{\varepsilon}_0$. The third term represents the effect of temperature, where $T*$ is defined as $T* = \frac{T - T_0}{T_m - T_0}$, where $T_m$ being the melting temperature, $T_0$ the reference temperature, and $T$ the current temperature. The constants $A, B, n$ and $m$ are obtained through experimental analysis.

### 2.2.4.2 Mie–Grüneisen equation of state model

The Mie–Grüneisen equation of state is used alongside the Johnson-Cook material model to describe the pressure-volume relationship of a solid at a given temperature. It provides two formulations depending on whether the material is in compression or expansion. The Grüneisen Γ is defined as:

$$\Gamma = \left(\frac{dp}{de}\right)_V \quad (2)$$

$$p - p_0 = \frac{\Gamma}{V}(e - e_0) \quad (3)$$

Where p_0, V and e_0 are pressure, volume, and internal energy at a reference state, typically the state where the temperature is assumed to be 0 K.

### 2.2.4.3 Johnson Cook dynamic failure model

Johnson Cook dynamic failure model describes the failure behavior of metals under high strain-rate loading. It determines damage based on the equivalent plastic strain at the element integration points. The damage parameter ω is expressed as,

$$\omega = \sum \left(\frac{\Delta \bar{\varepsilon}^{pl}}{\bar{\varepsilon}_f^{pl}}\right) \tag{4}$$

Where the plastic failure strain ($\bar{\varepsilon}_f^{pl}$) is given by

$$\bar{\varepsilon}_f^{pl} = \left[d1 + d2 \exp(d3^p_q)\right]\left[1 + d4\ \ln\left(\frac{\bar{\varepsilon}^{pl}}{\dot{\varepsilon}_0}\right)\right]\left[1 + d5\ \frac{(T - T_{room})}{(T_m - T_{room})}\right] \tag{5}$$

The first term in the Johnson-Cook failure model addresses the dependence of failure strain on pressure, while the second term accounts for strain rate effects. The third term captures the temperature dependence of failure. The constants d1, d2, d3, d4, and d5 are material-specific failure parameters that are experimentally determined. In a simulation, when the damage parameter ω reaches a value of 1 for an element, that element is deleted from the analysis. The Johnson-Cook model used in this study is available in the Abaqus/Explicit material library for both shell and solid elements.
In this study, a Hemispherical projectile is assigned in the initial condition of a softer material, particularly Aluminum A356. The Johnson-Cook Parameters for the same grade of Aluminum are taken from [52] and the detailed description of properties is listed in **Table A1.** Further analysis was then performed by considering a Harder Material, particularly AISI 4340 Steel. The Johnson cook parameters for the material are considered as per in [53] and a detailed description of the properties and parameters is tabulated in **Table A1.**

### 2.2.4.4 Ductile Damage Criteria

The link between stress and strain in the materials undergoing plastic deformation is the main focus of the ductile damage criteria. It indicates when the material experiences enough plastic strain to start failing. The plastic strain at failure, where stress increases until a certain point and then decreases as damage increases, is captured by the model. Practically speaking, as deterioration progresses, the material's ability to support loads decreases, causing a substantial displacement and, eventually, material collapse. Until the material achieves its maximal tensile strength and begins to fail, often seen as fracture during impact, events like bullet impacts or necking in tensile tests stress and strain are monitored. Predicting how and when the material will fail is made easier by the damage criterion. A Typical Ductile Damage Criteria along with the necessary parameters to be considered is shown in the Schematic Represented in **Fig 6**. Stress Triaxiality in a Simple Tensile Testing is given by (6),

$$Stress\ Triaxility, \eta = \frac{Hydrostatic\ (mean) Stress, P}{Equibvalent\ von\ Mises\ (deviatoric) Stress, \sigma_{eq}} \tag{6}$$

In the uniaxial Tensile Test, the stress tensor is given as (7),

$$S = \begin{bmatrix} \sigma_{11} & 0 & 0 \\ 0 & 0 & 0 \\ 0 & 0 & 0 \end{bmatrix} \quad (7)$$

Therefore, Hydrostatic (mean) Stress is given by (8),

$$P = tr(\mathbf{S}) = \frac{1}{3}(\sigma_{11} + \sigma_{22} + \sigma_{33}) = \frac{1}{3}\sigma_{11} \quad (8)$$

Also, equivalent von Mises (deviatoric) Stress is given as (9),

$$\sigma_{eq} = \sqrt{\frac{1}{2}[(\sigma_{11} - \sigma_{22})^2 + (\sigma_{22} - \sigma_{33})^2 + (\sigma_{33} - \sigma_{11})^2 + 6(\sigma_{12}^2 + \sigma_{23}^2 + \sigma_{31}^2)]} \quad (9)$$

$$\sigma_{eq} = \sigma_{11}$$

$$Stress\ Triaxility, \eta = \frac{P}{\sigma_{eq}} = \frac{\frac{1}{3}\sigma_{11}}{\sigma_{11}} = \frac{1}{3} \quad (10)$$

The Stress Triaxility for a uniaxial Tensile Test is given by as in (10)

Ductile Damage Criteria were applied to the impacted structure for which the properties of PMSCs were evaluated experimentally. Assignment of the properties for the overall structure was in gradation along the thickness of the structure as well as uniform material along the thickness representing the potential of each of the regions as discussed in **Section 2.2.3**. The Material Properties assignment for each of the regions is as tabulated in **Table A5**.

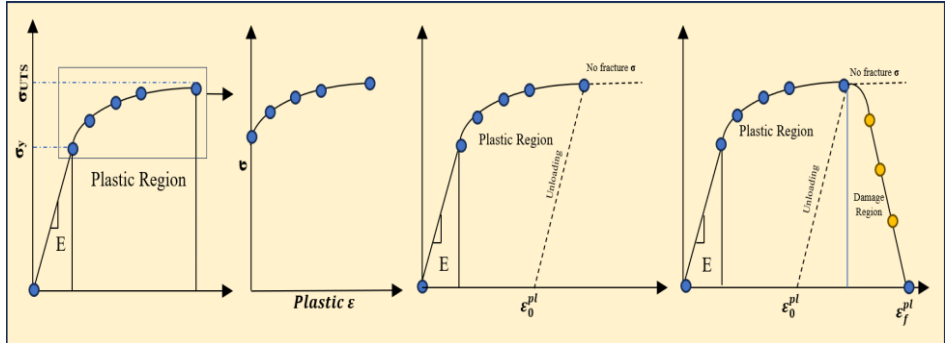

**Fig.6.** A Schematic of typical Ductile Damage Criteria with necessary parameters

## 3 Results and Discussion

### 3.1 Effective Properties Outputs

As discussed in **Section 2.1** the effective properties of RVE are extracted by imposing appropriate boundary conditions considering the spherical inclusion for which the Homogeneous isotropic properties are mentioned in **Table A3**. Subsequently, the matrix material properties are assigned as recorded in the manufacturer's catalog for LY 556 and HY 951 **Table A2**. A schematic of the different BCs applied on the RVE in order to extract the effective properties in respective directions and planes is shown in **Figure 7.** The effective capture of properties regarding RVE was done by varying the spherical inclusion sizes and volume fractions in the media.

The simulation study reveals that through this approach, a comprehensive understanding of how these parameters influence the critical material properties was developed.

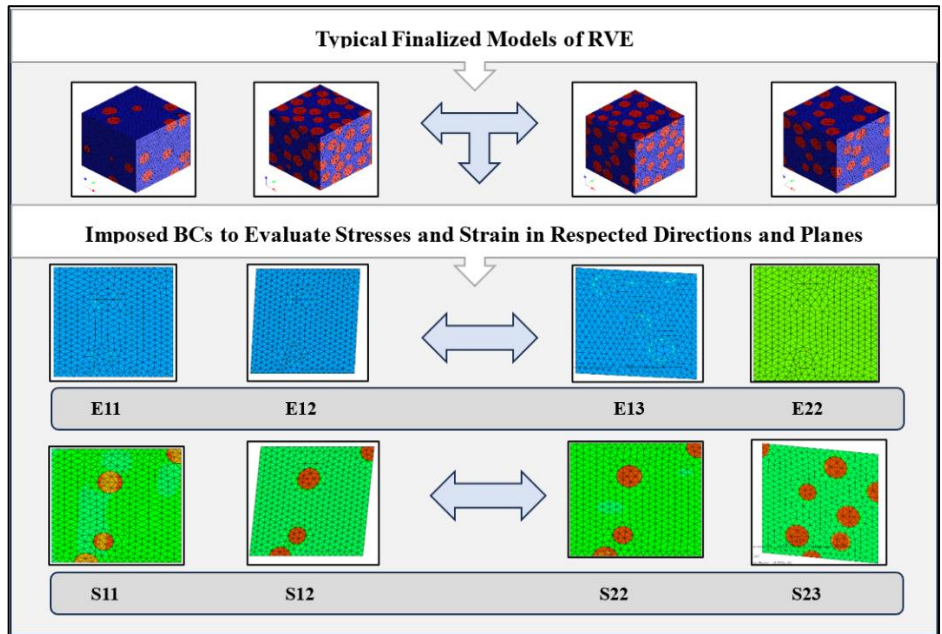

**Fig.7.** BCs applied to extract effective properties by varying inclusion size and volume fraction

The simulation results clearly demonstrate a significant increase in the elastic moduli in the three principal directions of the RVE i.e. $E_1$, $E_2$, and $E_3$, as both the inclusion diameter and volume fraction were increased as depicted in **Fig 8 a), b), c)**. This behavior can be attributed to the higher stiffness contribution from larger inclusions, which restrict the matrix deformation more effectively. As the inclusions become larger or more numerous, they act as rigid obstacles within the matrix, preventing excessive deformation under applied loads, thus leading to an increase in the overall stiffness of the composite material.

**Table A2:** Epoxy and Hardener cured properties (Manufacturer's Catalogue) [48]

| Properties | LY556 (Epoxy) and HY951/917 |
|---|---|
| Curing | 4 hours 80°C post-cure 8 hours 140°C |
| Viscosity | 10000-12000 mPa s |
| Shelf Life | 1.8 years |
| Modulus of Elasticity | 3100 – 3300 MPa |
| Density | 1.15 - 1.20 g/cm^3 |

**Table A3:** Sand Properties Derived from [47]

| Properties | Loose gravel high silica |
|---|---|
| Modulus of Elasticity (GPa) (Max) | 6.48 GPa |
| Poisson ratio | 0.2 – 0.35 |
| Density g/cm$^3$ | 1.2 – 1.52 |

Furthermore, the Poisson's ratios $\nu_{12}$, $\nu_{13}$, $\nu_{21}$, $\nu_{23}$, $\nu_{31}$, and $\nu_{32}$ showed a decreasing trend with the increase in inclusion size and volume fraction. This reduction can be reasoned through the fact that stiffer inclusions resist lateral deformations more effectively, leading to a smaller amount of transverse strain relative to the axial strain under load. Hence, the decrease in Poisson's ratios is consistent with the increase in the overall rigidity of the RVE.

In addition to the elastic moduli and Poisson's ratios, the study also observed a notable increase in the shear modulus $G_{12}$, $G_{23}$, and $G_{13}$ with increasing inclusion size and volume fraction as shown in **Fig 8. d), e), f)**. This enhancement in shear stiffness can be explained by the fact that larger inclusions provide better resistance to shear deformation, as the load is more evenly distributed and transferred through the rigid inclusions rather than being fully absorbed by the softer matrix. The matrix's ability to undergo shear is thereby limited, resulting in an overall increase in the shear modulus.

The material system, when evaluated as a whole, exhibited quasi-isotropic behavior, with minimal variation in material properties across different directions. This suggests that the composite system retains a uniform stiffness profile, closely resembling that of isotropic materials. This quasi-isotropic nature is especially beneficial in applications where uniform mechanical properties are desired in all loading directions, such as in structural or ballistic applications.

The further increase in volume fraction beyond 0.6 showed the incompatibility of packing of spherical inclusions in the RVE [58]. The further increment in volume fraction disrupts the 3D isotropy of mechanical properties [58]. This can be attributed [59] where the maximum dense random packing [59] $c_{DR}$=0.64. In conclusion, the study effectively captures the influence of inclusion size and volume fraction on the mechanical properties of the RVE. The results underline the critical role of these parameters in enhancing the stiffness and shear moduli while reducing lateral deformation, contributing to the material's overall robustness. This approach of varying inclusion geometry and distribution can serve as a powerful design tool in optimizing composite materials for high-performance applications, such as in lightweight, impact-resistant structures. The variation in the Inclusion size was from 0.1 to 1 mm in diameter and the volume fraction varied from 0.1 to 0.6 in most of the simulations carried out.

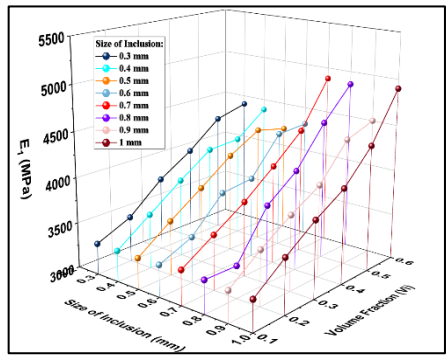

(a)

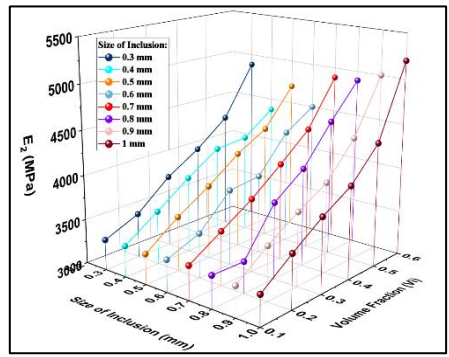

(b)

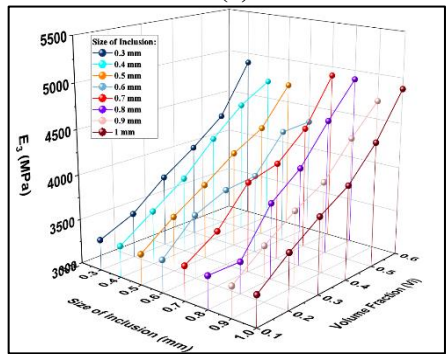

(c)

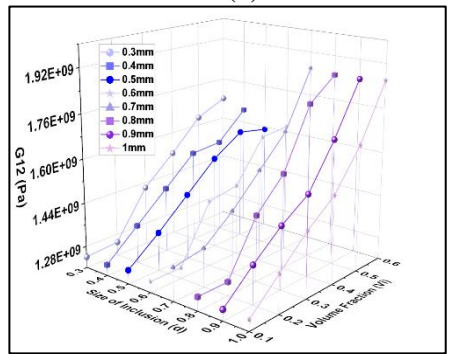

(d)

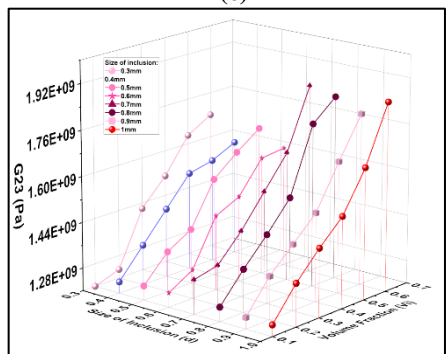

(e)

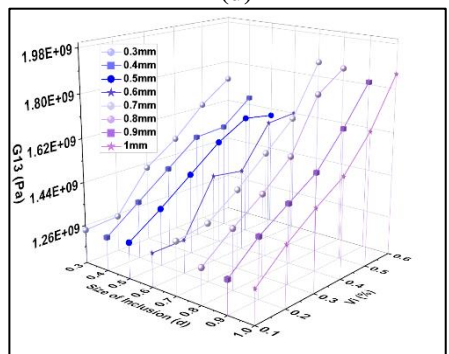

(f)

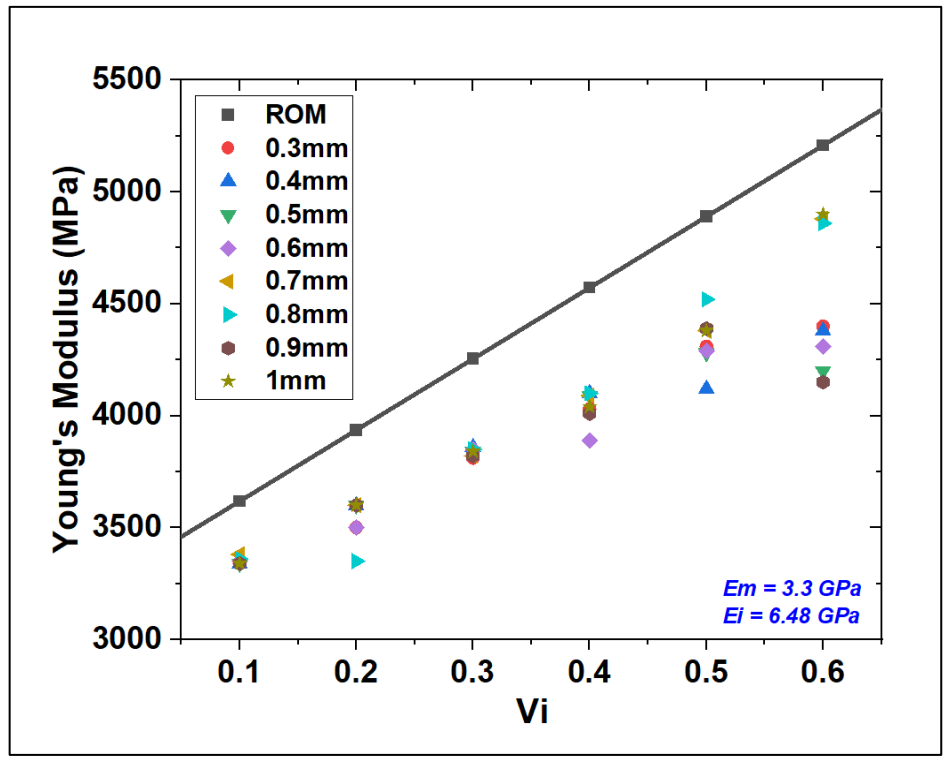

**(g)**
**Fig. 8. (a-f) Effective Properties**
**(g) ROM Compared with FEA outputs for different Sizes and Vi**

## 3.2 Computational Model Outputs

Computation Models with different configurations were considered which majorly involved two primary overall thicknesses i.e. 20mm and 40mm thick structures. Further in each of the thickness structures, computation models were created having unique material configurations within a single structure. Considering about 20mm thick structures, models consisted of 2 layers and 4 layers. The 2-layer configuration was further divided into 3 parts i.e. 2RG_20, 2RGU_20, and 2RGUE_20 for which the material configuration was utilized as in **Table A4.** The further structure for 20 mm thickness comprised of 4 equal parts for which the structures are particularly named as 4RG_20, 4RA_20. The 40mm thickness structures are further divided into 2 unique configurations for which the first set comprises structures having nomenclature i.e. 4RG_40, 4RGU_40, 4RGUE_40, 4RA_40 for which the material configurations were assigned utilized as in **Table A4.** The second set comprises configurations of 3RG_40 for which the material properties are tabulated in **Table A4.** The configuration for 2RG_20 consisted of two primary layers of equal thickness within a single 20 mm composite structure. The frontal layer was designed to be stiff and brittle, providing initial impact resistance, while the underlying layer was softer, contributing to energy absorption and structural support. 2RGU_20 comprised the same material property of the highest enhancement achieved by experimental testings throughout the 20mm thickness and 2RGUE_20 comprised Neat Matrix properties within the overall 20mm thickness of the structure. The 4-layered structure for 20 mm 4RG_20 comprised of 4 equally spaced layers with material properties varying from stiffest frontal to softest back layer in a single structure. The 4RA_20 configuration further has an alternate hard and soft layer of equal thickness within a single structure. For the 4RG_40 sample, 40mm thickness was divided into 4 equal layers with increased thickness, and a stiffest to softest gradation was incorporated in this structure for frontal to back layers. Similarly, 4RGU_40 comprised the stiffest configuration achieved through experimental terms to the whole structure, and 4RGUE_40 comprised nea matrix properties as given in **Table A4.** The 4RA_40 comprised alternate harder and softer regions with increased thickness. A 40 mm thick structure 3RG_40 consisted of 3 graded regions for which the gradation was from stiffer to softer material as given in **Table A4.**

All the structural configurations were subjected to extensive simulation studies across a range of projectile velocities to determine the ballistic performance specific to each configuration. The schematic of each configuration is specified in **Fig 9**. These studies facilitated the identification of the ballistic limit defined as the maximum projectile velocity at which the structure can resist full perforation under impact. This parameter is crucial for assessing the ballistic resistance and structural robustness of the composite designs. The projectiles were modeled using Johnson-Cook (JC) steel material properties, as detailed in **Table A1.** and elaborated in **Section 2.2.4**.

The overall projectile dimensions adhered to the specifications provided in **Section 2.2.1**, ensuring consistency in impact analysis. Boundary conditions were strategically applied as outlined in **Section 2.2.2**, where axisymmetric boundary conditions were leveraged to optimize computational efficiency while minimizing resource demands. This approach enabled an accurate representation of symmetric impact responses without incurring excessive computational costs.

The comprehensive simulations captured critical velocities for each configuration, highlighting the maximum energy absorption capacity and failure thresholds. The layered models included Ductile Damage Criteria Ellaboarated in **Section 2.2.4.4** and Material Configurations in **Table A5**. Such rigorous modeling and analysis provide a thorough understanding of the dynamic interactions between projectiles and composite structures, forming the basis for advanced impact mitigation strategies.

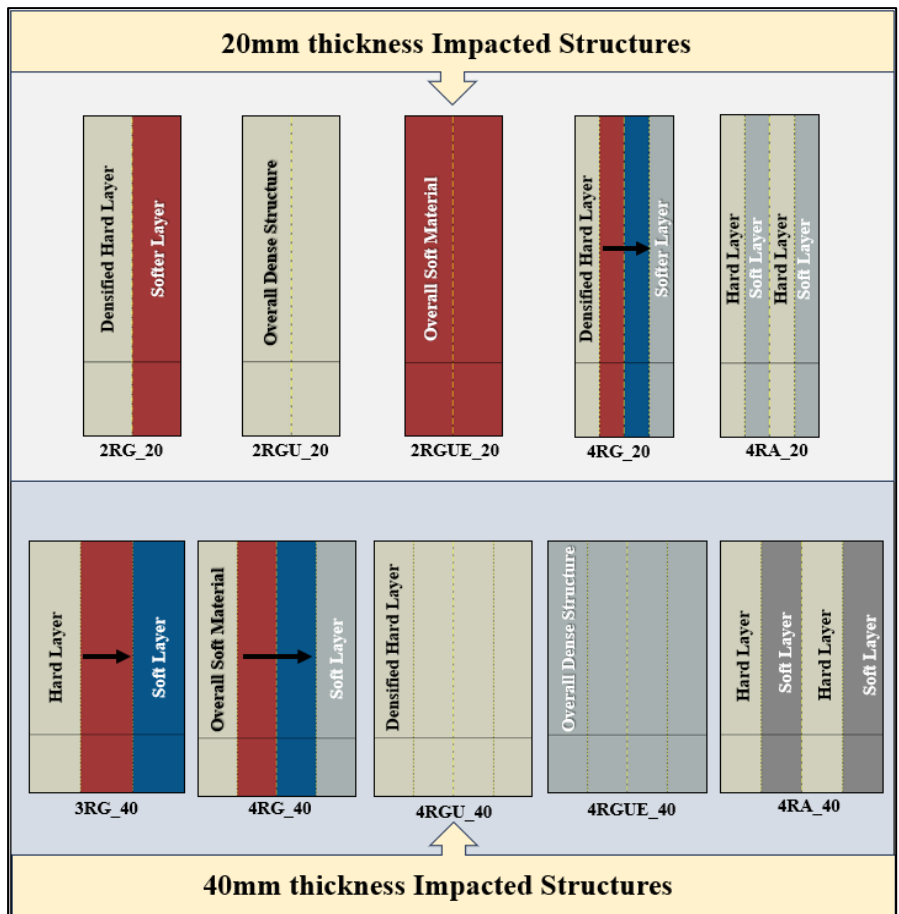

**Fig 9.** Schematic of Impacted Structures configuration utilized in computational studies

This enables a comprehensive understanding of how energy is absorbed, dissipated, and transferred through the material upon projectile impact. Additionally, the percentage of kinetic energy retained by the bullet after impact was quantified to determine the proportion of energy that is not dissipated by the material but rather remains with the projectile. This metric is essential for evaluating the material's ability to resist perforation and its energy absorption capacity, especially in the context of multi-layered composite systems.

Depth of penetration, defined as the distance a projectile travels into the material upon impact at a given velocity, was extensively studied for each configuration. This parameter serves as a key indicator of the ballistic resistance of the composite, reflecting the material's ability to resist penetration under varying projectile velocities. A comprehensive study of penetration depth was conducted across different configurations to evaluate the structural effectiveness in resisting projectile intrusion. Each configuration's resistance to penetration was tested under a range of velocities to determine the ballistic performance across various impact conditions.

Moreover, in certain configurations, the phenomenon of partial perforation with rebound was observed, where the bullet partially penetrated the structure but bounced back or was retained within a certain distance from the surface. This behavior is indicative of the material's energy dissipation efficiency and its ability to prevent full penetration while mitigating the bullet's momentum. The occurrence of this rebound effect under specific velocities is crucial for assessing the ability of the structure to absorb and redistribute the kinetic energy of the projectile, thus enhancing its multi-hit capabilities and resilience under successive impacts. This rebound phenomenon is a critical consideration in the design of ballistic-resistant structures, as it highlights the material's capacity to mitigate projectile damage without full perforation.

The simulation studies incorporated various material configurations to analyze the extent of penetration resistance and the performance of individual layers in response to different impact scenarios. These findings contribute to the development of more robust, multi-layered composite systems that are optimized for ballistic protection, offering a combination of energy dissipation, resistance to perforation, and enhanced performance at different projectile velocities.

### 3.3     Performance Evaluation of 20mm Composite Structures

### 3.3.1 Kinetic Energy of Bullet Vs Time Studies for Several Configurations

The structures of 20 mm thickness were initially analyzed for ballistic impact studies, with each configuration varying in the number of layers and the distinct material properties allocated to those layers. These properties were assigned based on experimental outcomes and corresponded to the unique configurational properties detailed in **Table A4**. The structural configurations featured gradations through the thickness, as discussed in **Section 3.3**. Detailed schematics of these structural arrangements are illustrated in **Fig .9.** The impact simulations involved the use of a hemispherical steel projectile, with specifications and properties provided in **Section 2.2** and **Table A1**. The simulations focused on tracking the Kinetic Energy (K.E.) loss over time, emphasizing the percentage of kinetic energy dissipated by the projectile as it penetrated through the layered structures with different material configurations. These studies offered insights into how variations in the structural arrangement influenced the energy absorption and ballistic resistance of the composite structures, highlighting the effectiveness of each configuration in mitigating the impact force and preventing complete perforation.

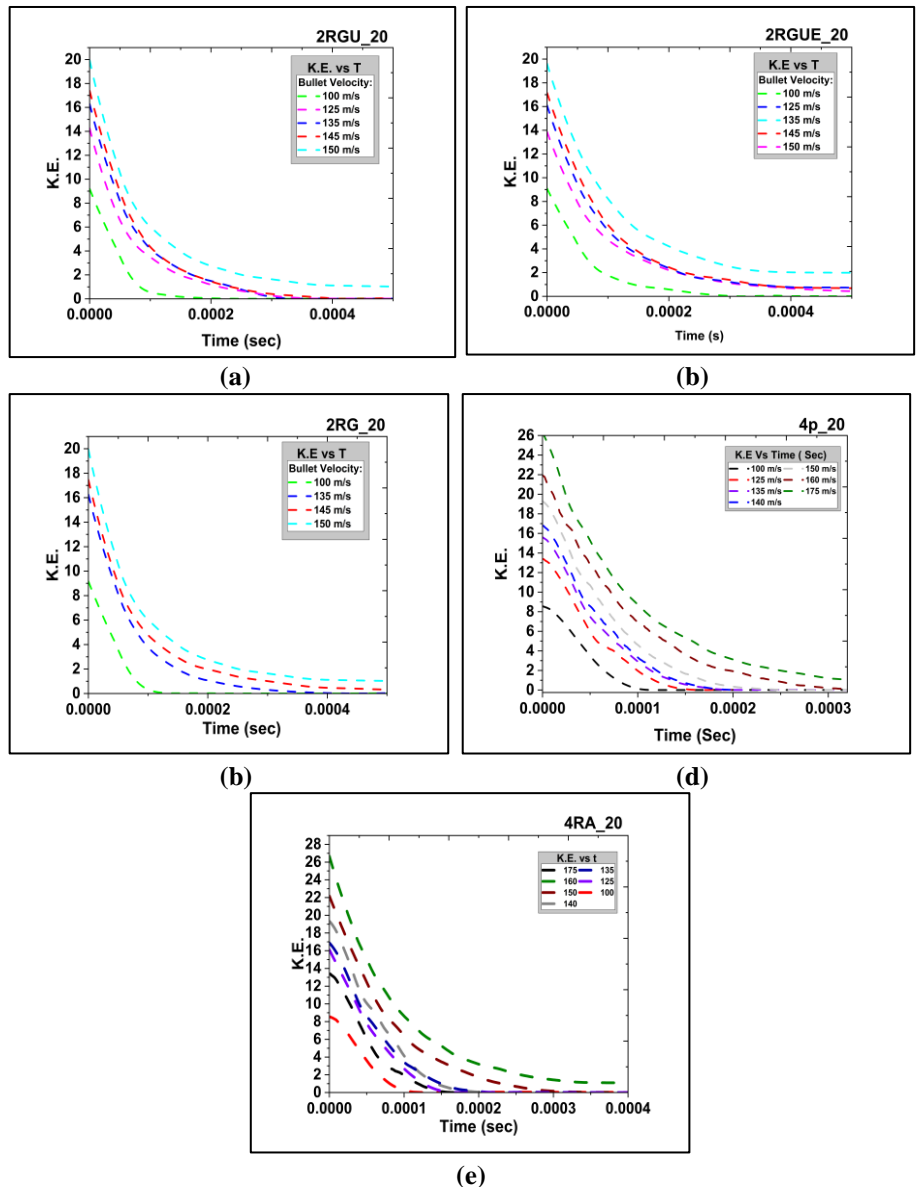

**Fig 10.** Kinetic energy vs Time for 20mm varied structural configurations

In **Fig .10.** the kinetic energy (K.E.) of the bullet over time is illustrated for each structural configuration evaluated during the simulation studies. The projectile velocity was varied incrementally from 100 m/s to 150 m/s to effectively capture the ballistic limit for each structural arrangement. Particular observations were made with the 2RGUE_20 configuration, which represents a composite structure composed entirely of a neat matrix material. The kinetic energy profile for this structure indicates a gradual decrease over time, suggesting a relatively slow dissipation rate of the bullet's kinetic energy as it interacts with the composite **Fig.10 b)**. This behavior is indicative of limited resistance from the material, allowing for easier penetration by the projectile.

The underlying reason for the 2RGUE_20 structure's reduced resistance lies in the inherent properties of the neat matrix material, which lacks significant reinforcement [12]. Unlike reinforced composites or structures with graded density incorporating harder inclusions, the neat matrix material is comparatively soft and ductile, with lower hardness and tensile strength [13]. Consequently, as the bullet maintains its momentum, the material's relatively poor hardness and lack of reinforcing agents prevent effective opposition to penetration, causing full perforation at higher impact velocities and an inability to adequately dissipate kinetic energy over time [12]. It lacks mechanisms such as brittle fracture, erosion of the projectile tip, or energy reflection that is often observed in stiffer or more complex multi-phase composites.

The structure designated as 2RGU_20, characterized by its densely graded composition and superior mechanical properties, exhibited the highest resistance to projectile impact among the evaluated configurations. This structure's enhanced performance can be attributed to its advanced material characteristics, including peak tensile strength, modulus of elasticity, hardness, and impact strength [4]. The grading within the 2RGU_20 configuration was designed to optimize energy absorption and stress distribution during high-velocity impacts. Upon analyzing the kinetic energy (K.E.) dissipation profile, it was observed that the K.E. of the projectile consistently approached zero when impacted at velocities up to 145 m/s **Fig. 10 a)**. This indicates that the structure was capable of absorbing significant kinetic energy, effectively arresting the projectile before full perforation occurred. The combination of high modulus and tensile strength facilitated the formation of stress concentrations that could propagate controlled, localized deformation, effectively dissipating energy and resisting crack propagation [5]. The uniform distribution of high-strength material within the graded layers contributed to a composite response where the initial impact was mitigated through both elastic and plastic deformations [13]. The material's elevated hardness played a crucial role in resisting surface indentation and initial penetration, while the high impact strength ensured that the structure could absorb the dynamic load without rapid failure or delamination [8].

At a projectile velocity of 150 m/s, the structure exhibited full perforation, delineating a ballistic limit of approximately 148 m/s for the 2RGU_20 configuration. This threshold is indicative of the maximum impact velocity the structure can endure while preventing complete penetration. The exceptional energy absorption and redistribution properties, coupled with its graded material profile, make 2RGU_20 one of the most effective configurations for high-velocity impact resistance within the studied models. The structure layered design facilitated a progressive failure mechanism, wherein each layer contributed incrementally to reducing the projectile's kinetic energy and delaying complete structural failure.

While the performance benefits of the 2RGU_20 structure are evident, the higher weight becomes a crucial concern, especially for applications where weight reduction is critical. The weight increase could result in higher load-bearing requirements and may impact the efficiency or maneuverability of the structure in practical scenarios.

To address this issue, a promising approach is to optimize the material gradation through the incorporation of multiple layers with varying properties [23-24]. By using a graded structure, it is possible to replicate the performance characteristics of the densely graded 2RGU_20 structure while reducing overall weight. The gradation allows for the material properties to transition from stiff to soft, thereby distributing the load more effectively and reducing the need for high-density material throughout the entire structure. This strategy not only maintains ballistic resistance but also ensures a lighter, more efficient structure, balancing performance with weight reduction **Fig. 10 c)**.

Studies were conducted on the 4RG_20 and 4RA_20 structures, which feature gradation variations from denser to softer layers, and alternate dense and soft layers, respectively. The 4RG_20 structure employs a gradual transition from the stiffest frontal layer to the softest rear layer, while the 4RA_20 structure alternates between dense and soft layers across its thickness. These configurations were analyzed to assess their impact on ballistic resistance and energy dissipation, with a focus on escalating performance while managing weight and material distribution for defense applications.

The 4RG_20 structure, when impacted, showed a gradual reduction in kinetic energy **Fig. 10 d)**, displaying a consistent energy dissipation profile similar to that of the 4RA_20 structure **Fig. 10 e)**. Both structures exhibited a gradual, but steady decline in kinetic energy, indicating their ability to effectively dissipate the projectile's energy during penetration. However, the full perforation occurred at a projectile velocity of 175 m/s for both configurations, with a slight variation in the ballistic limits: 171 m/s for the 4RG_20 structure and 169 m/s for the 4RA_20 structure. This marginal difference in ballistic limit can be attributed to the structural gradation and material distribution. While the 4RG_20's uniform gradation from stiff to soft layers allows for a more controlled energy transfer across the structure, the 4RA_20's alternate layering introduces a stepwise energy dissipation, leading to slightly less effective energy absorption in certain regions [56]. Furthermore, both configurations demonstrated a reduction in weight compared to a fully dense structure, with the 4RG_20 structure offering a more uniform distribution of material properties, thereby enhancing its ballistic resistance without significantly increasing mass. The weight reduction, coupled with their comparable ballistic limits, positions both structures as viable candidates for defense applications, where a balance of weight and ballistic resistance is crucial. By optimizing layer gradation, these structures offer an effective means of improving performance while mitigating the adverse effects of excessive weight.

### 3.3.2 Percentage K.E. loss vs Depth of Penetration Studies for several structural configurations

The figures in **Fig. 11.** illustrate the percentage kinetic energy of the projectile left while penetrating to the depth of the structure which shows the energy dissipation estimate for different configurations with the maximum or minimum perforation impacted upon the same or different velocities.

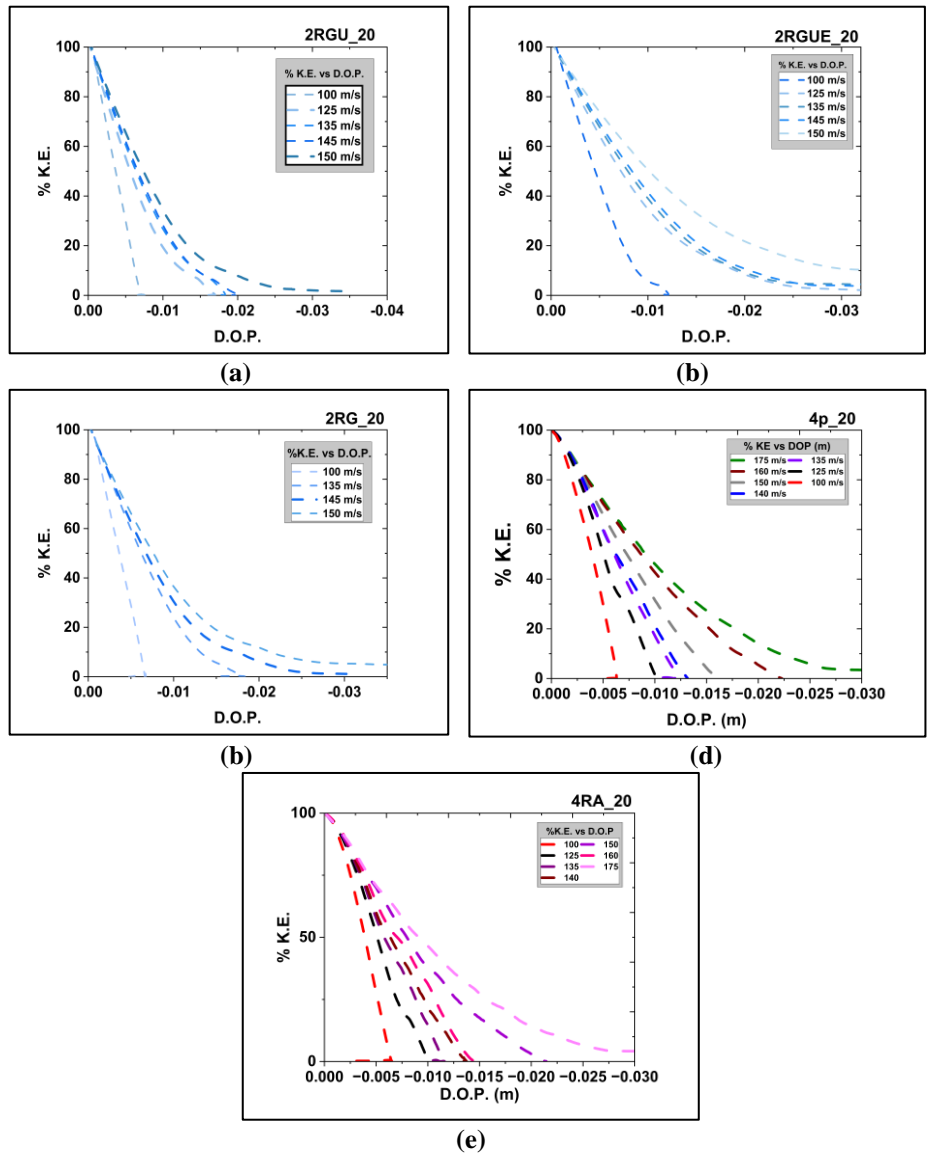

**Fig 11.** Kinetic energy (%K.E.) vs Depth of penetration (D.O.P)

In the 2RGUE_20 structure, characterized by neat matrix properties, kinetic energy (K.E.) loss measurements revealed a significant dissipation of energy, resulting in a depth of penetration (D.O.P.) of approximately 12 mm **Fig. 11 b)**. This led to the effective arrest of the projectile, halting its movement completely and demonstrating partial rebound due to residual elastic resistance within the material. The structure's performance at velocities exceeding 125 m/s showed full penetration, with kinetic energy reductions recorded at 98%, 96%, 95%, and 90% for projectile velocities up to 150 m/s. This trend underscores the limitation of the neat matrix configuration, where lower stiffness and strength parameters could not sufficiently impede higher-velocity impacts, thereby allowing greater projectile penetration and energy transmission.

Comparatively, the 2RGU_20 structure, which integrates a highly densified gradient of reinforcement, demonstrated superior energy absorption capabilities. For a 100 m/s impact, the D.O.P. was contained at 7 mm, with full kinetic energy capture and a subsequent slight rebound, indicating enhanced resistance to penetration due to a higher modulus and impact strength **Fig. 11 a)**. The response to higher velocities (125, 135, and 145 m/s) showed incremental D.O.P. values of 18, 19, and 20 mm, respectively, with significant energy absorption and retention, indicating that the densified matrix provided robust localized stress distribution, facilitating delaying crack propagation. The 2RGU_20 structure's complete perforation at 150 m/s resulted in a substantial 98% energy dissipation, highlighting its ballistic threshold.

For the 2RG_20 graded structure, kinetic energy retention and penetration depth were analyzed, showing results close to those of the 2RGU_20 configuration **Fig. 11 c)**. The energy dissipation across graded layers suggested an intermediate balance of resistance, where the transition from stiff to more compliant material sections contributed to gradual energy absorption. The distributed response of the graded structure helped manage stress waves effectively, though not match the full kinetic capture of the densified 2RGU_20 structure. These results indicate that the compositional gradation and reinforcement distribution significantly impact ballistic resistance, influencing energy dissipation, crack propagation, and structural deformation.

The structural configurations 4RG_20 and 4RA_20 demonstrated comparable K.E. absorption profiles as the projectile penetrated through the layers, indicating an increase in ballistic performance compared to prior configurations. The 4RG_20 configuration achieved approximately 75% K.E. absorption over three distinct layers, resulting in a cumulative D.O.P. of 15 mm for a projectile velocity of 171 m/s. The remaining 25% of K.E **Fig. 11 d)**. was dissipated in the final layers, leading to a D.O.P. of 19.5 mm, with minimal energy resistance observed in the last graded section. Similarly, the 4RA_20 structure displayed a marginally different response, with a 19.7 mm D.O.P. at a slightly lower ballistic limit of 168 m/s.

These results indicate that both structures effectively utilize layered gradation to manage stress wave propagation and energy dissipation. The alternating dense and soft layering of 4RA_20 facilitated a distributed deformation mechanism, preventing rapid fracture propagation, while the gradually graded 4RG_20 structure maintained continuous energy absorption, enhancing the overall resistance against perforation **Fig. 11 e)**. *The findings suggest that layer composition and arrangement significantly influence the ballistic limit and the structural integrity under high-velocity impacts.*

### 3.3.3 Depth of Penetration with Rebound Characterization of Different Configurations

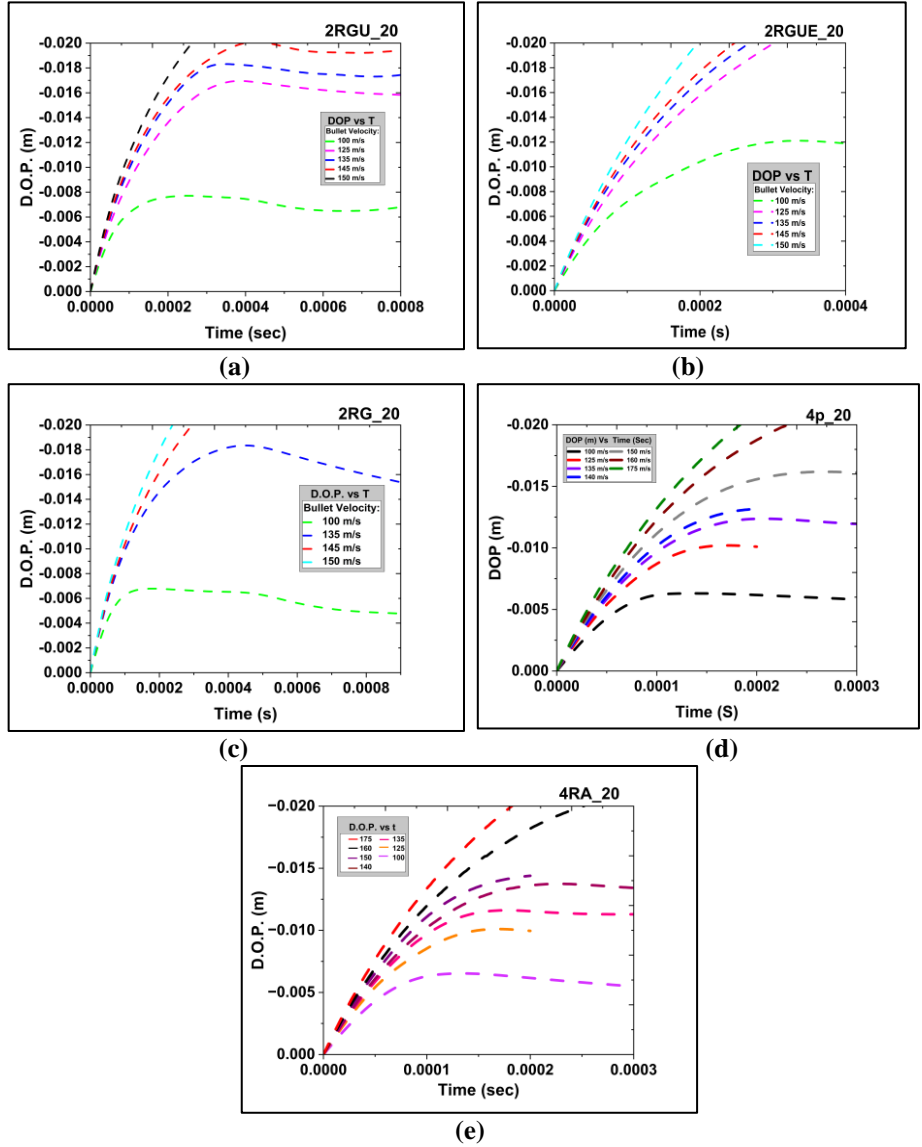

(a) (b)

(c) (d)

(e)

**Fig 12 .** D.O.P vs Time studies for different 20mm configurations

The D.O.P. versus time analysis for different structural configurations highlighted distinct rebound behaviors in **Fig.12.** The 2RGU_20 structure, with its uniformly distributed material properties, exhibited a lower rebound compared to the 2RG_20 graded structure at the same impact velocities of 135 m/s and 100 m/s. This is due to the uniform composition's ability to distribute and absorb energy evenly, minimizing elastic recovery.

In contrast, the graded 2RG_20 structure, transitioning from stiffer to softer layers, showed higher rebound as the softer rear layers released stored energy more effectively, contributing to elastic recovery. The 4RA_20 configuration, with alternating dense and soft layers, also demonstrated a higher rebound than 4RG_20. The abrupt changes between dense and compliant layers in 4RA_20 led to uneven energy storage and rapid release, enhancing rebound upon impact.

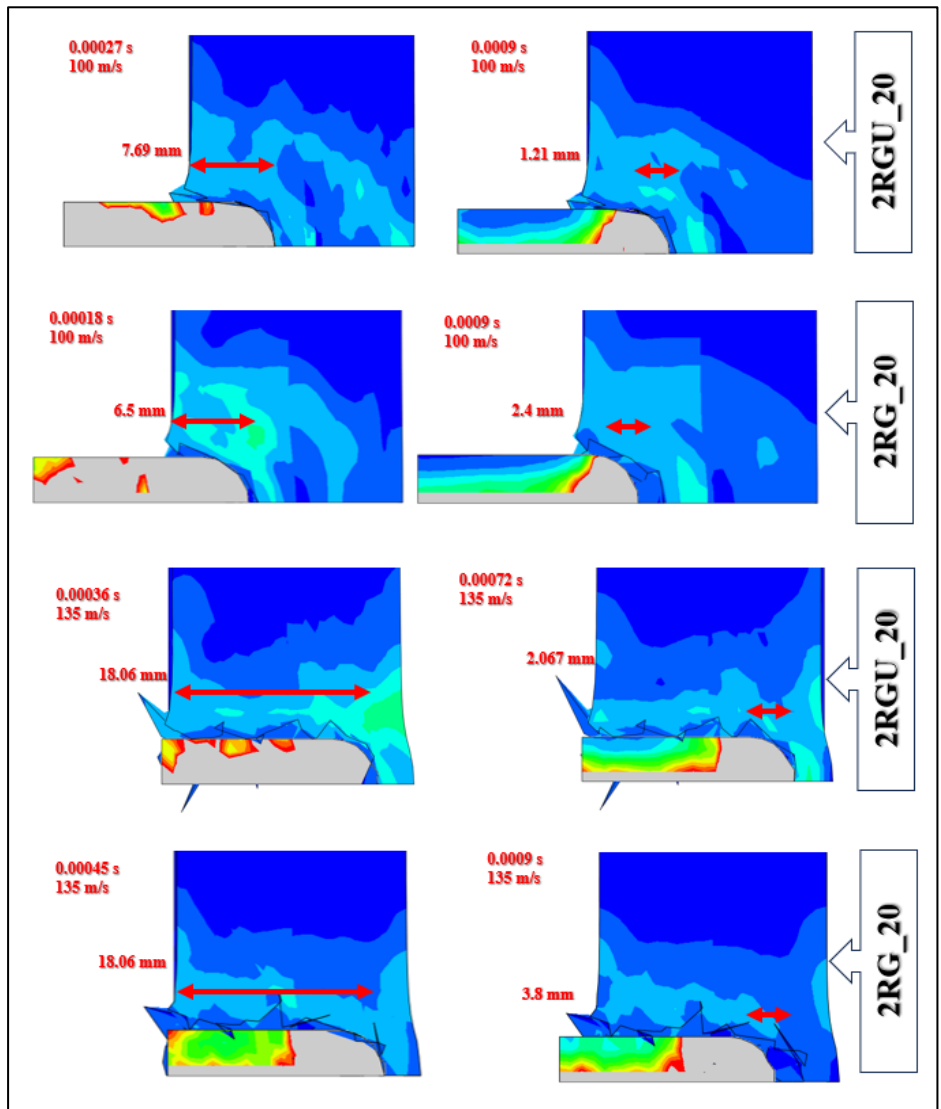

**Fig 13.** Rebound upon varying projectile velocities for 2RG_20 and 2RG_20

The **Fig 13.** represents the simulation studies carried out for the 2RGU_20 and 2RG_20 configurations when impacted at two different velocities, 100 m/s and 135 m/s. It was observed that the projectile penetrated to a depth of 7.69 mm and had a rebound of 1.21 mm in the 2RGU_20, which is a uniform, densified structure. In the 2RG_20, when impacted at the same 100 m/s velocity, the projectile penetrated up to 6.5 mm with a rebound of 2.4 mm. When the projectile velocity was further increased to 135 m/s and impacted on the 2RGU_20 configuration, the D.O.P. was observed at a maximum of 18.06 mm with a rebound of 2.067 mm. Lastly, when the same velocity was imparted on the 2RG_20 configuration, the results showed a rebound of 3.8 mm.

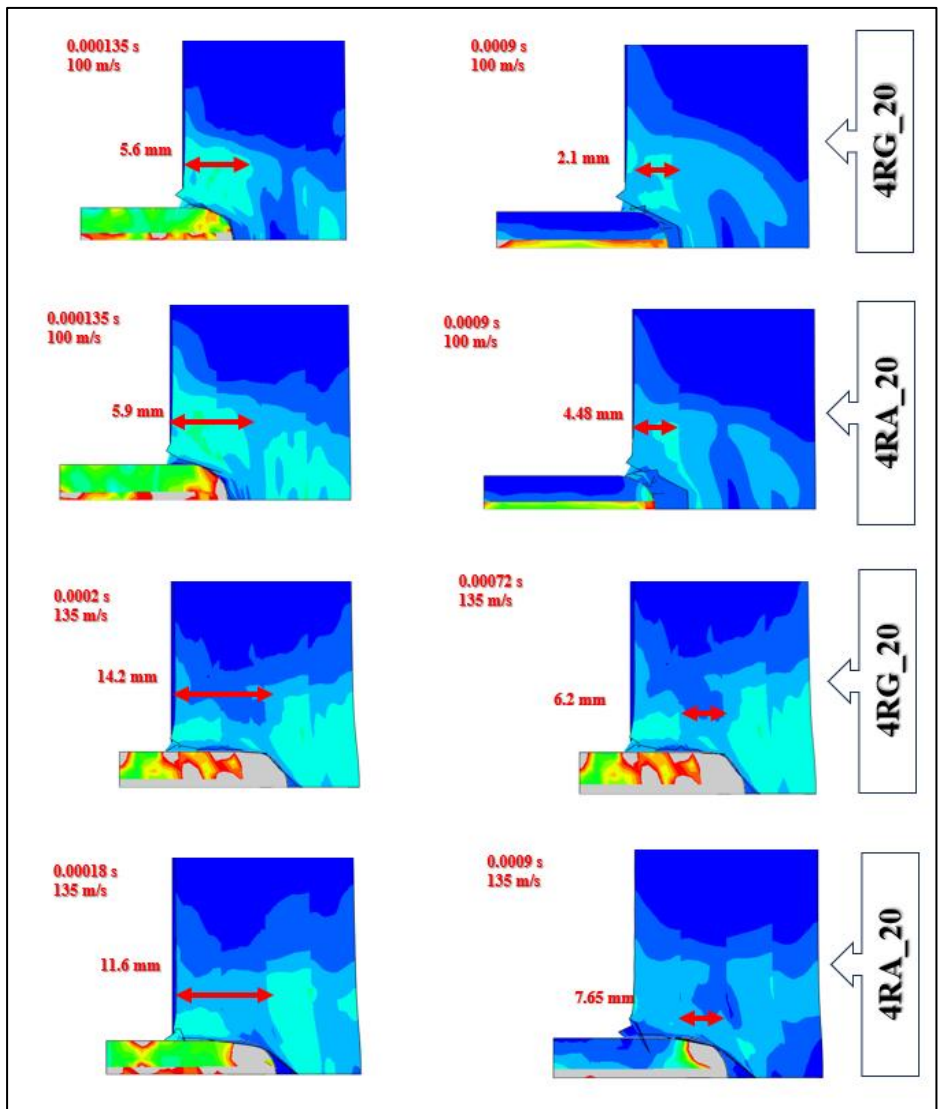

**Fig 14**. Rebound upon varying projectile velocities for 4RG_20 and 4RA_20

The higher rebound in graded structures like 2RG_20 is due to the mix of stiff and softer layers. The softer layers recover more elastically after impact, increasing rebound. In contrast, the uniform dense structure 2RGU_20 has consistent stiffness, absorbing energy more evenly and limiting rebound. With increased projectile velocity, *the rebound effect is more pronounced, especially in graded materials, as they have varied responses from different layers that contribute to higher elastic recovery.*

Similarly, the trend continued for the 4RG_20 and 4RA_20 structures under impacts of 100 m/s and 135 m/s as shown in **Fig 14.** . The bullet penetrated to a depth of 5.6 mm with a rebound of 2.123 mm for the 4RG_20 structure under 100 m/s impact, while the 4RA_20 structure showed a D.O.P. of 5.9 mm and a rebound of 4.481 mm. When the velocity was increased to 140 m/s, the rebound rose to 6.2 mm for 4RG_20 and 7.65 mm for 4RA_20. This demonstrates that while the initial penetration depths were comparable, the graded structures, especially those alternating between dense and soft layers like 4RA_20, exhibited greater elastic recovery, leading to higher rebound.

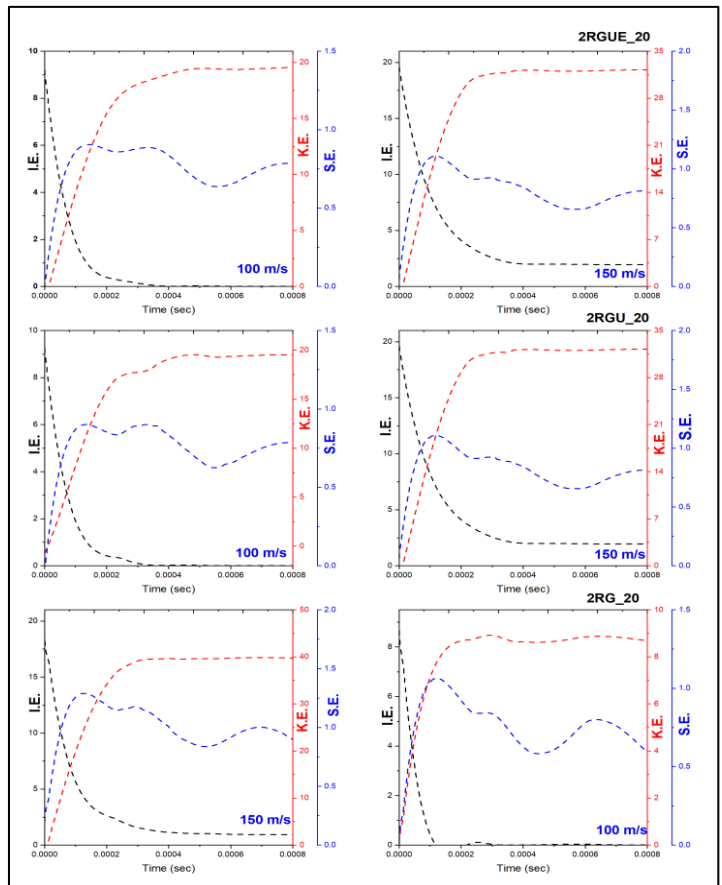

**(a)**

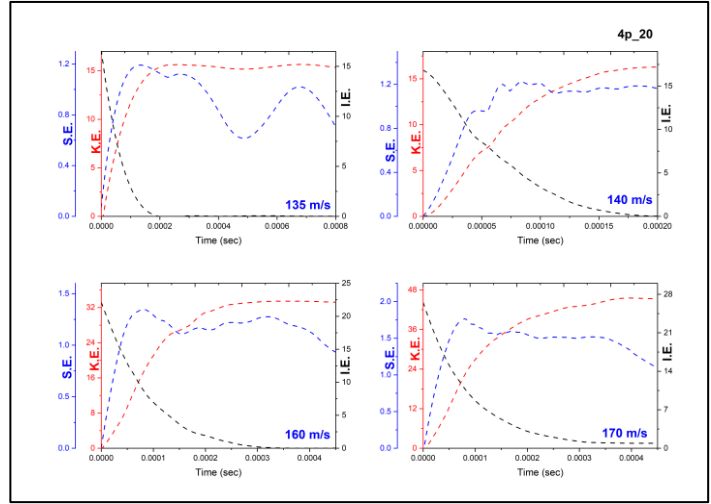

**(b)**

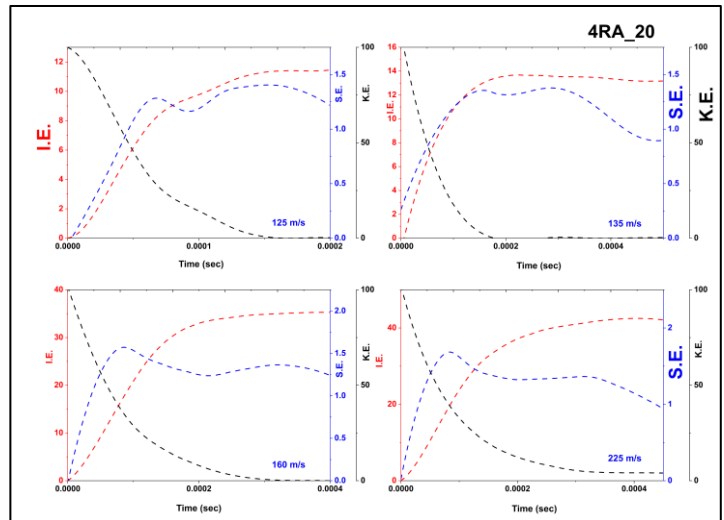

**(c)**

**Fig 15.** I.E., S.E., K.E. vs Time for different structural configurations

The **Fig 15.** depicting changes in Internal Energy, Strain Energy, and K.E. over time provides insights into the energy dynamics during ballistic impact. Internal Energy is the energy absorbed by the structure, including elastic strain energy, energy dissipated through inelastic processes like plasticity, and any artificial strain energy introduced during simulation [57]. Strain Energy represents the energy stored within the material as it deforms, while K.E. shows the reduction in the projectile's energy as it penetrates the material. Together, these energy components reflect the material's ability to resist impact, absorb energy, and manage energy dissipation, crucial for evaluating the structure's ballistic performance.

### 3.4 Performance Evaluation of 40mm Composite Structures

### 3.4.1 K.E. of Bullet Vs Time Studies for Several Configurations

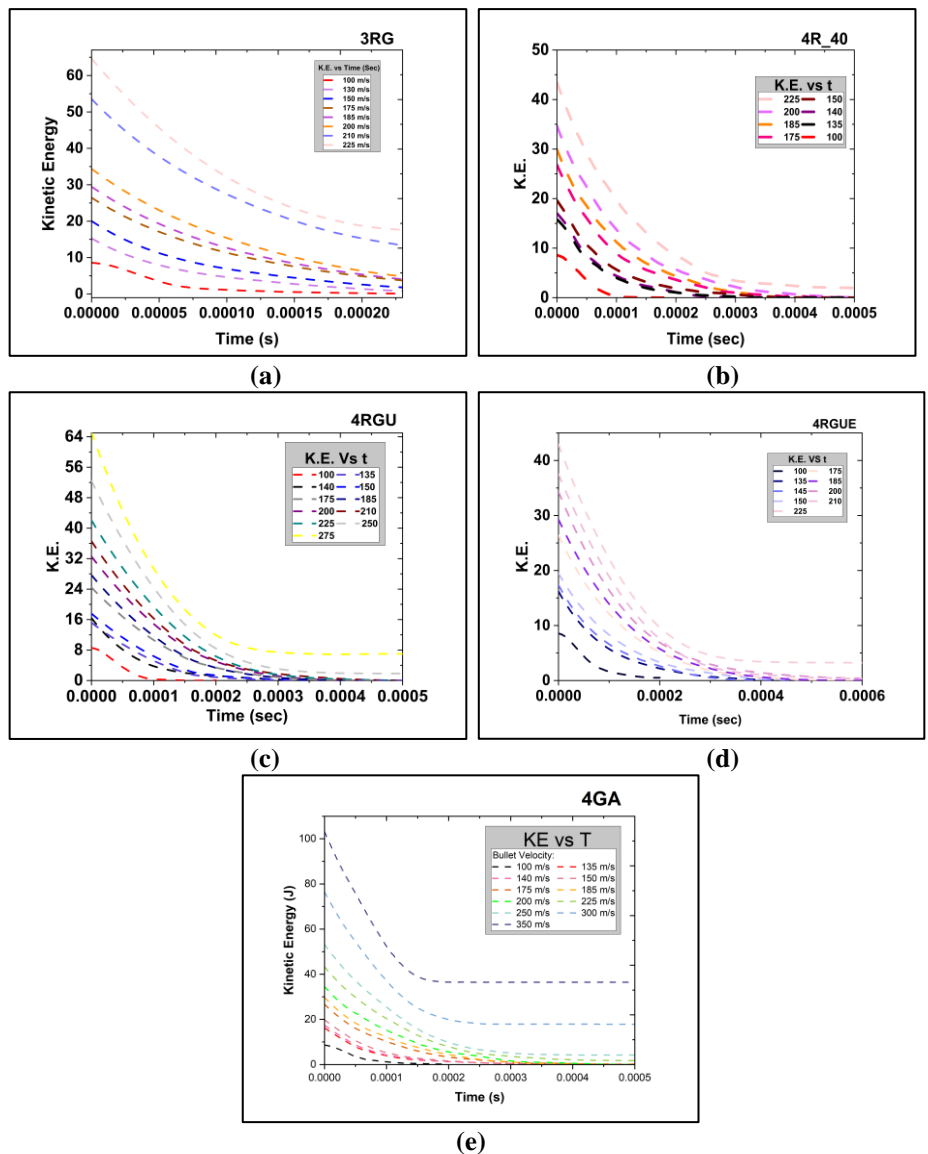

**Fig 16.** Kinetic energy vs Time for 40 mm varied structural configurations

The **Fig 16.** presents the K.E. vs. Time profiles for the various structural configurations utilized in the simulation studies, with each configuration's designation outlined in **Section 3.1.2**. The study focuses on four distinct structural designs, each varying in layer configurations and material properties. These include a 3-layer structure, 4-layer structure, uniform densified structure, and uniform neat matrix property structure. These configurations were impacted by projectiles at different velocities, to analyze how each design performs under varying ballistic conditions. By evaluating the changes in kinetic energy over time, the study offers insights into the energy dissipation characteristics of each structure, shedding light on their respective abilities to absorb and resist the impact energy, thus providing a comprehensive comparison of their performance across multiple scenarios. A study was conducted to investigate the effect of increasing layers, as seen in the previous structures, on the ballistic performance of the materials. **Fig.16 a)** displays the 3RG_40 configuration, where the kinetic energy K.E. of the projectile progressively decreases towards zero as the impact velocity reaches approximately 200 m/s, indicating a comparable yet slightly improved performance when compared to the 4RGUE_40 structure, which consists of a uniform neat matrix property. The K.E. for 4RGUE_40 structure tends towards zero till 185 m/s showing a marked difference in the ballastic limit **Fig.16 d)**. This suggests that the incorporation of gradation in the 3-layer configuration enhances its ability to absorb and dissipate energy more efficiently than a structure composed solely of a neat matrix material. The observed trend continues as the study moves to the 4-layer structure, as discussed in **Section 4.3**, where the impact of gradation becomes more pronounced. The inclusion of gradation significantly influences the K.E. reduction, highlighting the effectiveness of layering in improving the ballistic resistance of the structure. This result emphasizes the potential of gradation to optimize energy absorption and enhance the overall ballistic performance of composite materials, especially when considering varying impact velocities. In **Fig.16 b)** and **Fig.16 c)** the graded and the uniform densification structures show the extent to which they can sustain perforation. The trend continues with the ballistic limit reaching the 4RG_40 structure at about 225 m/s and the 2RGU_40 structure at about 235 m/s. The comparable difference is negligible concerning weight as discussed in **Section 2.1. Fig.16 e)** represents the structure 4RA_40 impacted by various velocities with similar outcomes as in 20mm thickness results showing a reduced ballistic limit than the previous two discussed.

### 3.4.2 Percentage K.E. loss vs Depth of Penetration Studies for several structural configurations

The Figures in **Fig 17.** illustrate the percentage of kinetic energy %K.E. retained by the projectile as it penetrates through the depth of structures with varying configurations. For the 40 mm thick structures, the trend shows an increase in energy absorption similar to that observed in the 20 mm thick structures. However, the 40 mm structures demonstrate a greater loss in kinetic energy, indicating enhanced energy dissipation capabilities as the projectile pierces deeper. This behavior suggests that thicker graded and uniformly densified structures contribute to improved ballistic resistance by effectively reducing the kinetic energy of the projectile during penetration.

**Fig. 17 a)** shows the plot for the 3RG_40 structure with 3 layers graded from stiffer to softer gradation along the thickness in 3 layers. The characteristics curves show that the bullet loses most of the K.E. within half of its depth up to 150 m/s further beyond this up to 200 m/s projectile reaches nearly the extremity of the structure showing the ballastic limit of the configuration. Further full perforation is seen with about 70% K.E. reduction for 250 m/s impact and 65 % K.E. loss for 275 m/s impact. **Fig.17 b)** and **Fig. 17 c)** show the plots for 4RG_40 and 4RGU_40 respectively. 4RGU_40 structure effective reduction of the K.E. is observed while maintaining the D.O.P. The velocity up to 175 m/s was effectively captured till the depth of 20 mm and till 225 m/s didn't show full perforation. The ballistic limit with effective capture of Bullet's K.E. was observed to be 235 m/s for this structure. Further, the effective capture of 95 % and 90 % of K.E. was seen at velocities of 250 and 275 m/s respectively.

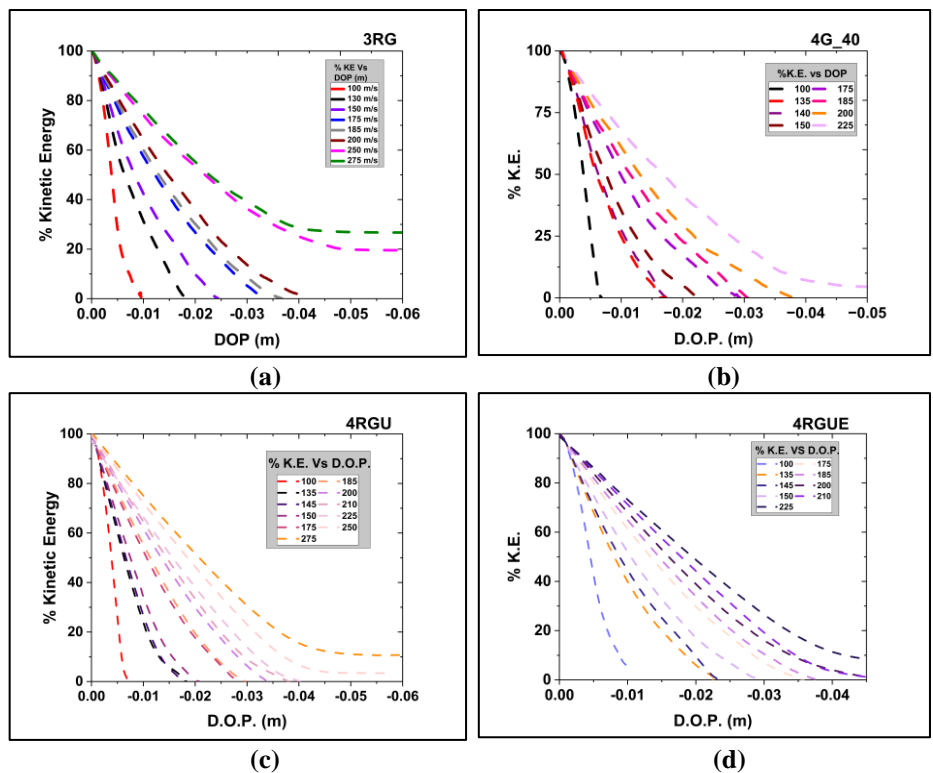

(a)    (b)

(c)    (d)

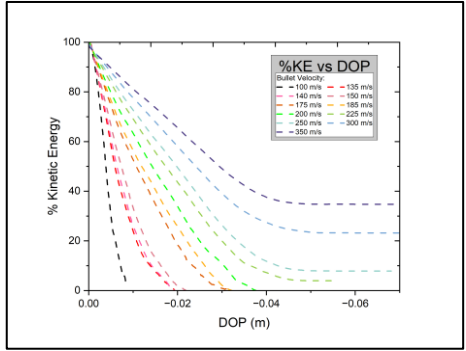

**(e)**

**Fig17 .** %Kinetic energy (%K.E.) vs Depth of penetration (D.O.P)

In the 4RG_40 structure, a marginal variation was observed with a comparable densified structure. The structure showed the ballistic limit of 250 m/s with further capturing the K.E. for the above velocities in an effective manner and the results align with the previously discussed results for 20 mm thickness graded structures.

**3.4.3 D.O.P. with blackface deformation studies for varied configurations**

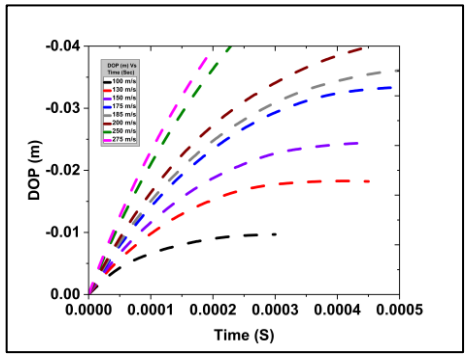 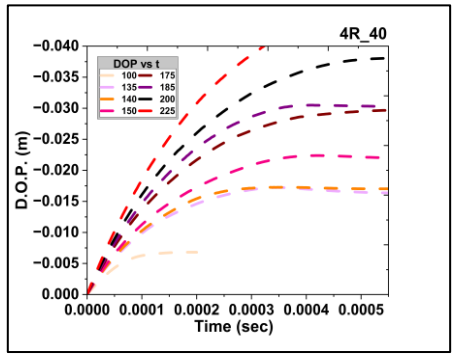

**(a)**  **(b)**

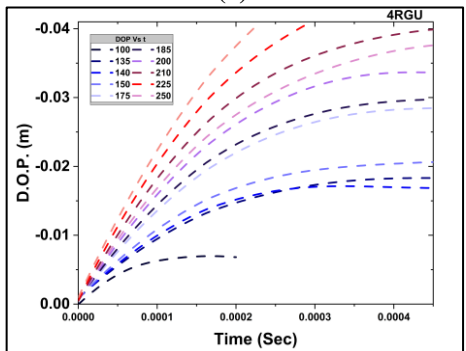 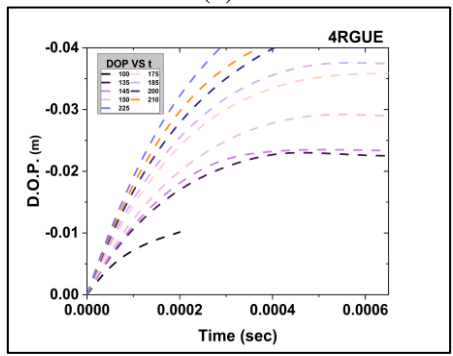

**(c)**  **(d)**

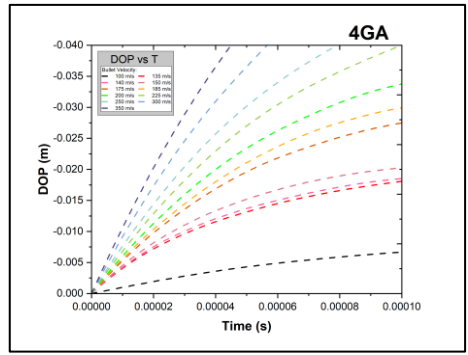

**(c)**

**Fig 18.** D.O.P vs Time studies for different 40mm configurations

The Figures in **Fig 20.** show D.O.P. of different configurations of the ballistic structure for 40 mm thickness when impacted with incremental velocities. The D.O.P. studies provided a significant understanding of the influence of gradation in a single structure and how exactly it affects projectile penetration. In **Fig. 20 a)** which is of 3RG_40 structural configuration the structure captured the resistance to penetration in the frontal stiffer layer for most of the velocities.

The structure adhered to the Ballistic limit of 200 m/s for this configuration showing a slight increase in the critical velocity than the 4RGUE_40 which was 185 m/s as revealed in **Fig. 20 d)**. The highest D.O.P for the simulations carried out for 3RG_40 was 35 mm for 185 m/s with the D.O.P. touching the thickness for about 200 m/s and in 4RGUE_40 the height D.O.P. was about 37 mm for 185 m/s velocity. Further, when 4RG_40 and 4RGU_40 were tested structures showed maximum D.O.P. of 37 mm for first for velocity of 200 m/s and 39 mm for 210 m/s **Fig.20 b), c)**. The 4GA_40 structural configuration showed a maximum D.O.P. according to simulation studies of about 32 mm 200 m/s showing a better performance due to the thicker stiffer layer in the front resisting the projectile penetration in the initial layer **Fig. 20 e)**.

The Simulations for 40 mm thickness structures thus revealed that *the Graded structures provide an increment in resistance to projectile impact with a marginal variation in ballistic limit for uniform structures contributing to the strength-to-weight ratio of overall structures.*

Further, the Backface deformation was observed for 40 mm thickness structures considering the maximum velocity to which the maximum backface deformation is observed. In **Fig 19.** the maximum travel of the projectile in a structure along with the maximum deformation for the particular velocity is represented. For the 3RG_40 structure when impacted with 200 m/s a D.O.P. of 42.2 mm was observed with 6.18 mm backface deformation showing a maximum amongst all the other's observed. The 4RG_40 and 4RGU_40 structures when impacted by 200 m/s and 225 m/s respectively the D.O.P. of 37.9 and 40 mm was observed with a reduced Backface signature of 3.95 mm and 4.91 mm. Further when 4RGUE_20 and 4RA_40 Structures were impacted with the velocities of 185 m/s and 200 m/s the D.O.P. observed was 37 mm and 37.5 mm showing a decreased Backface signature of 1.9 mm and 2.5 mm which was enhanced when compared with all of the structures observed.

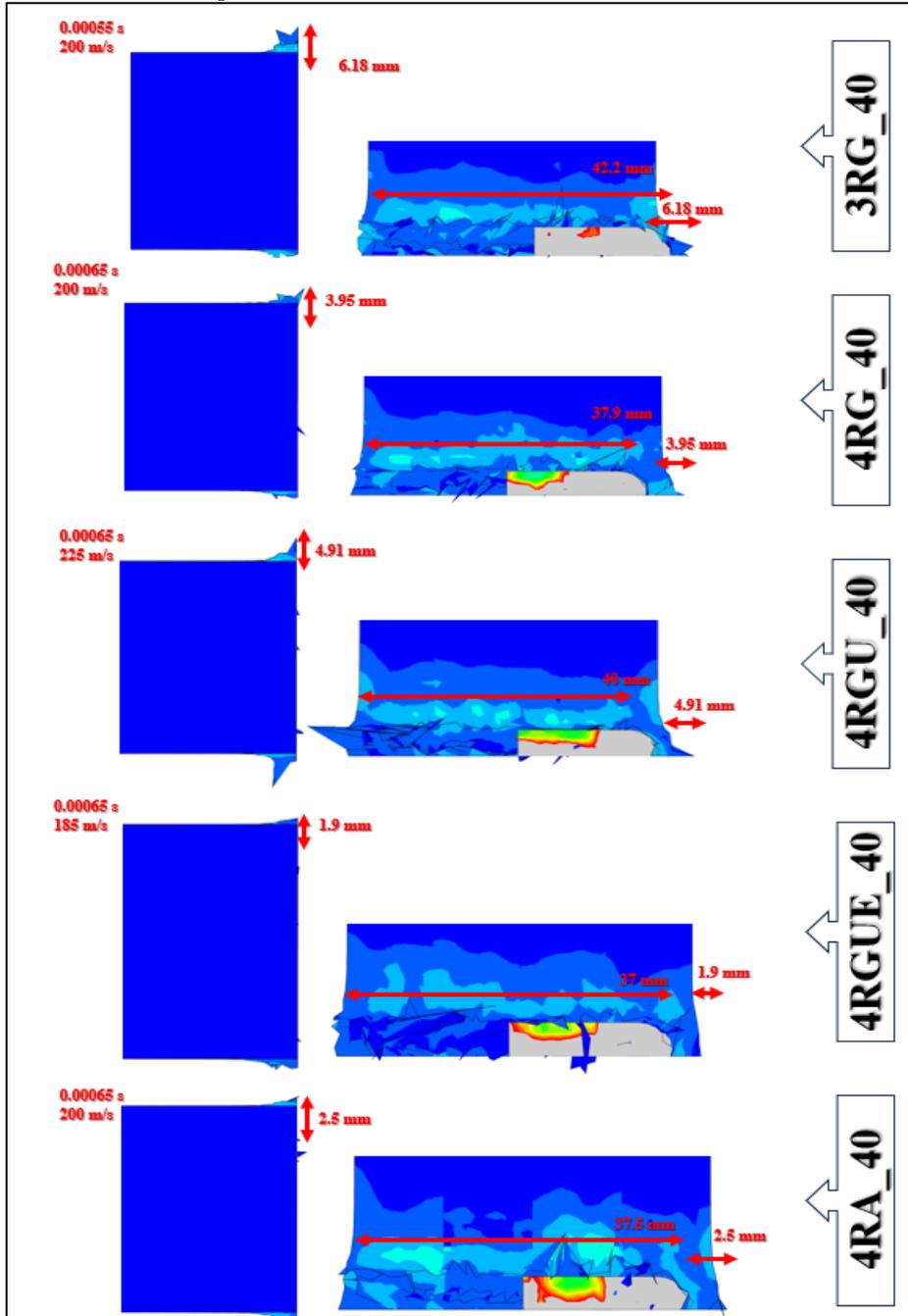

**Fig 19.** 40mm Structures with Max Backface Deformation upon Particular Velocities

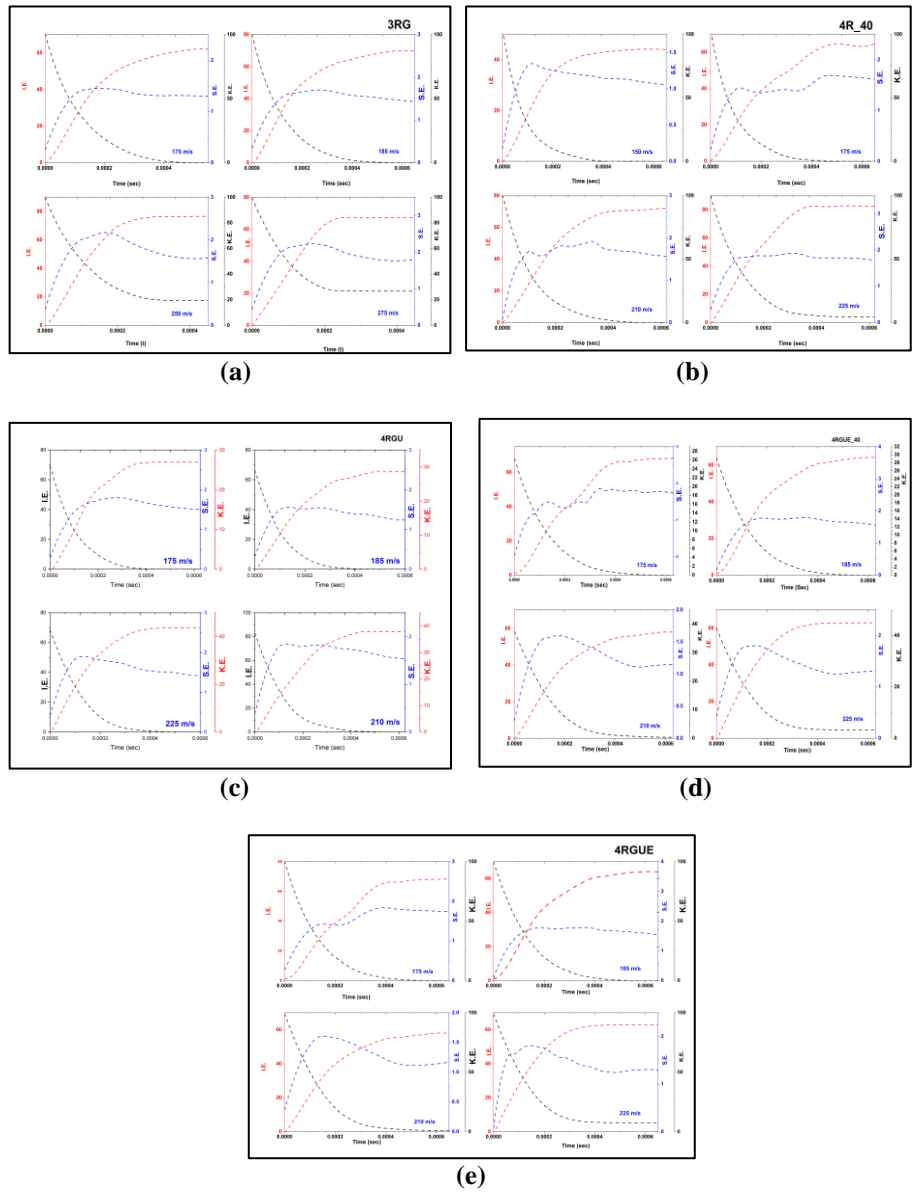

**Fig 20.** I.E., S.E., K.E. vs Time for different 40 mm structural configurations

The investigation indicated that *maximum backface deformation occurred in graded structures with fewer layers*, which resulted in a less efficient load transfer across the material interfaces. The transition of stresses was less seamless, leading to pronounced localized deformation. Conversely, *graded structures with a greater number of layers exhibited enhanced performance*, as the kinetic energy of the projectile was more effectively absorbed within a shallower depth of penetration (D.O.P.), particularly at higher impact velocities. This translated to reduced backface deformation and superior momentum absorption, indicative of efficient stress distribution and energy dissipation throughout the material. The uniform structure with neat matrix properties demonstrated the lowest backface signature at lower velocities, highlighting the capacity of softer matrices to undergo substantial plastic deformation and dissipate energy efficiently [56]. Configurations with alternately graded layers, particularly those with increased thickness and higher layer count, showed improved performance, characterized by reduced backface deformation and a higher ballistic limit, emphasizing their capability for superior impact resistance and energy management.

# 4     Conclusion

In this work, a thorough computational analysis complemented by a preliminary experimental investigation has been carried out for the novel Polymer matrix sand Functional graded composites to extract the effective stacking sequence for enhanced ballistic impact resistance. The previous study involved the characterization of PMSCs to extract optimal properties with an improvising fabrication process for variation in inclusion size and volume faction for different polymer matrices. The Experimental Properties extracted in the previous study are utilized in the current study to design an effective stacking sequence by combining different material properties adhered layers within a single structure with an objective of effective strength-to-weight ratio maintained for the structure. Some of the critical observations revealed are as stated.

1. The computational studies carried out for 20 mm thickness structures showed an enhanced ballistic limit for the Uniformly densified structures with a negligible difference for the 4 layered stiff to soft graded structures showing an effectiveness in strength to weight ratio for the graded media **Section 3.3**.
2. Structures transitioning from a stiff to a soft layer with only 2 layers demonstrate a lower ballistic limit compared to 4-layered configurations. This is primarily due to the more abrupt change in mechanical properties within the 2-layer system, leading to higher stress concentrations and inefficient energy dissipation upon impact. In contrast, 4-layered structures provide a more gradual transition in mechanical properties, allowing for better energy absorption and distribution throughout the material. 4RGU_20 configuration showed 171 m/s ballistic limit and 4RG_40 showed 168 m/s ballistic limit with reduced weight showing the potential of gradation with minimal variation **Section 3.3.1**.
3. Additionally when studied with alternate hard and soft gradations the enhancement in ballistic limit was not observed compared to the graded structures for the same number of layers the ballistic limit was stagnant for 20 mm structures **Section 3.3.3**.
4. The projectile was seen to be rebounded by the resistance of the structure for which the graded structures provided a larger rebound than the uniform structures and less layered structures. Increased velocities showed a higher rebound with up to 6.2 mm and 7.48 mm for 4RG_40 and 4RA_40 structures when impacted with 140 m/s projectile velocity **Section 3.4**.

5. The trend continued to the 40 mm studies as well with a spiked enhancement in ballastic limit for graded as well as densified uniform structures. The D.O.P. studied revealed that most of the Bullet's K.E. was dissipated within the initial impact and further initial layers.
6. With a notable improvement in K.E. dissipation Graded and uniform structures continued to further capture the Bullet's K.E. for velocities beyond the Ballistic limit with a range in between 75 % to 98 % K.E. capture. This showed the effectiveness of layered structures.
7. The Backface deformation studies revealed that with decreased layers there was a significant backface deformation for particular velocities limiting the structure's efficiency. With an increase in layer number a lesser backface signature was observed with a higher projectile velocities impacted. Further, the Alternate hard and soft gradation showed the least backface deformation impacted upon the similar velocities due to most of the K.E. absorbed in thicker Stiffer layers and the cushioning effect provided by softer layers of the structure **Section 3.4.3**.

The thus analyzed Polymer matrix sand inclusion Functionally Graded composites showed a viable enhancement in ballistic impact resistance. Providing an optimal thickness and layered configurations with lightweightness and cost-effectiveness as a prominent factor aligned with these structures. Future work aims to improve the optimality of these Functionally Graded structures by integrating diverse inclusions, including hyperplastic materials, to further enhance their ballistic performance within a single structure.

**Appendices (A)**
See Tables A.1-A.6

**Table A.1**
**Material Properties the Johnson-Cook (JC) material the model used in the Computational model derived from [53]**

| Parameter | Description | AISI 4340 |
|---|---|---|
| D1 | Failure Parameter | 0.05 |
| D2 | Failure Parameter | 3.44 |
| D3 | Failure Parameter | -2.12 |
| D4 | Failure Parameter | 0.002 |
| D5 | Failure Parameter | 0.61 |
| $\rho$ (kg·m-3) | Density | 7830 |
| E (GPa) | Young's Modulus | 205 |
| A (MPa) | Yield Stress | 792 |
| B (MPa) | Hardening Constant | 510 |
| n | Hardening Exponent | 0.26 |
| c | Strain Rate Constant | 0.014 |
| m | Temperature Softening Exponent | 1.03 |
| Tm (K) | Melting Temperature | 1519.9 |
| $\epsilon$ (s^-1) | Reference Strain Rate | 1 |

**Table A.2**
**Material Properties for respective Epoxy grades Derived from [54] & [48]**

| Property | Epoxy - Trans ER099 | Epoxy - LY556 |
|---|---|---|
| Viscosity (at 25 degree) | 450-650 mPa s | 10000-12000 mPa s |
| Shelf Life | 2 years | 1.8 years |
| Modulus of Elasticity | 7 x 10^6 kg/mm^2 | 3100 - 3300 MPa |
| Density | 1.13 g/cm^3 | 1.15 - 1.20 g/cm^3 |
| Epoxy Content | 5.5 - 5.7 | 5.3 - 5.45 |

**Table A.3**
**Sand Properties Derived From [47]**

| Properties | Loose gravel high silica |
|---|---|
| **Modulus of Elasticity (GPa) Max** | 6.48 GPa |
| **Poisson ratio** | 0.2 – 0.35 |
| **Density** | 1.2 – 1.52 |

**Table A.4**
**Samples configuration for simulation impact studies**

| | 2RG_20 | 2RGU_20 | 4RA_20 |
|---|---|---|---|
| 20 mm thickness | Layer 1- 70% of 0.3mm sand and Layer 2 Neat Epoxy | Overall - 70% of 0.3mm sand size | Layer -1,3 - 70% 0.3mm sand<br>Layer 2,4 - 20% 0.3mm sand. |
| | 2RGUE_20 | 4RG_20 | |
| | Overall - Neat Epoxy | Layer 1 - 70% 0.3mm sand<br>Layer 2- 20% 0.425mm sand<br>Layer 3- 20% 0.3mm sand.<br>Layer 4- Neat epoxy | |
| | 3RG_40 | 4RG_40 | 4RA_40 |
| 40 mm thickness | Layer 1 - 70% 0.3mm sand<br>Layer 2- 20% 0.425mm sand<br>Layer 3- Neat epoxy | Layer 1- Neat epoxy<br>Layer 2- 20% 0.3mm sand.<br>Layer 3- 20% 0.425mm sand<br>Layer 4- 30% 0.425mm sand | Layer -1,3 - 70% 0.3mm sand<br>Layer 2,4 - 20% 0.3mm sand. |
| | 4RGU_40 | 4RGUE_40 | |
| | Overall - 70% of 0.3mm sand size | Overall - 70% of 0.3mm sand size. | |

**Table A.5**
**Ductile damage parameters**

| $w_i$ | Size | E(Pa) | μ | Yield Stress (Pa) | Plastic Strain | Stress Triaxiality | Displacement at Failure |
|---|---|---|---|---|---|---|---|
| 0.2 | 0.425 | 1.21E+09 | 0.35 | 2.01E+07 | 0.019 | 0.33 | 2.30E-03 |
| | | | | 2.15E+07 | 0.02 | | |
| 0.3 | 0.425 | 1.31E+09 | 0.35 | 2.59E+07 | 0.029 | 0.33 | 3.40E-03 |
| | | | | 2.67E+07 | 0.03 | | |
| 0.4 | 0.425 | 1.33E+09 | 0.35 | 3.16E+07 | 0.0255 | 0.33 | 3.20E-03 |
| | | | | 3.29E+07 | 0.0275 | | |
| 0.5 | 0.425 | 1.38E+09 | 0.35 | 3.67E+07 | 0.03 | 0.33 | 3.50E-03 |
| | | | | 3.77E+07 | 0.031 | | |
| 0.6 | 0.425 | 1.49E+09 | 0.35 | 4.05E+07 | 0.032 | 0.33 | 3.80E-03 |
| | | | | 4.15E+07 | 0.034 | | |
| 0.7 | 0.425 | 1.11E+09 | 0.35 | 1.27E+07 | 0.0139 | 0.33 | 1.60E-03 |
| | | | | 1.29E+07 | 0.014 | | |
| 0.2 | 0.3 | 1.20E+09 | 0.35 | 3.15E+07 | 0.03235 | 0.33 | 3.55E-03 |
| | | | | 3.27E+07 | 0.0327 | | |
| 0.3 | 0.3 | 1.45E+09 | 0.35 | 3.65E+07 | 0.032 | 0.33 | 3.55E-03 |
| | | | | 3.71E+07 | 0.037 | | |
| 0.4 | 0.3 | 1.59E+09 | 0.35 | 3.69E+07 | 0.029 | 0.33 | 3.40E-03 |
| | | | | 3.79E+07 | 0.03 | | |
| 0.6 | 0.3 | 1.65E+09 | 0.35 | 3.57E+07 | 0.025 | 0.33 | 2.94E-03 |
| | | | | 3.67E+07 | 0.0259 | | |
| 0.7 | 0.3 | 1.81E+09 | 0.35 | 3.91E+06 | 0.0257 | 0.33 | 3.45E-03 |
| | | | | 4.10E+06 | 0.0259 | | |
| Neat | | 1.19E+09 | 0.35 | 3.12E+07 | 0.029 | 0.33 | 3.40E-03 |
| | | | | 3.22E+07 | 0.03 | | |